\documentclass[twocolumn,superscriptaddress,showpacs,amsmath,floatfix,notitlepage]{revtex4-1}
\usepackage{amsmath}
\usepackage{amssymb}
\usepackage{bm}
\usepackage{epsfig}
\usepackage{graphicx}
\usepackage{color}

\newcount\timehh  \newcount\timemm
\timehh=\time \divide\timehh by 60
\timemm=\time
\begin{document}
\title{High resolution study of the yellow excitons in Cu$_2$O subject to an electric field}

\author{J. Heck\"otter}
\affiliation{Experimentelle Physik 2, Technische Universit\"{a}t
Dortmund, D-44221 Dortmund, Germany}
\author{M. Freitag}
\affiliation{Experimentelle Physik 2, Technische Universit\"{a}t
Dortmund, D-44221 Dortmund, Germany}
\author{D. Fr\"ohlich}
\affiliation{Experimentelle Physik 2, Technische Universit\"{a}t
Dortmund, D-44221 Dortmund, Germany}
\author{M. A\ss mann}
\affiliation{Experimentelle Physik 2, Technische Universit\"{a}t
Dortmund, D-44221 Dortmund, Germany}
\author{M. Bayer}
\affiliation{Experimentelle Physik 2, Technische Universit\"{a}t
Dortmund, D-44221 Dortmund, Germany}
\affiliation{Ioffe Institute,
Russian Academy of Sciences, 194021, St.-Petersburg, Russia}

\author{M. A. Semina}
\affiliation{Ioffe Institute, Russian Academy of Sciences, 194021,
St.-Petersburg, Russia}
\author{M. M. Glazov}
\affiliation{Ioffe Institute, Russian Academy of Sciences, 194021,
St.-Petersburg, Russia}

\begin{abstract}
We have used high resolution transmission spectroscopy to study the exciton level spectrum in Cu$_2$O subject to a longitudinal external electric field, i.e., in the geometry where the transmitted light is propagating along the field direction. Different experimental configurations given by the field orientation relative to the crystal and the light polarization have been explored. We focus on the range of small principal quantum numbers $n \leq 7$. The number of exciton states belonging to a particular principal quantum number increases with $n$, leading to an enhanced complexity of the spectra. Still, in particular for $n = 3 \ldots 5$ a spectral separation of the different lines is feasible and identification  as well as assignment of the dominant state character are possible. We find a strong dependence of the spectra on the chosen light propagation direction and polarization configuration, reflecting the inadequacy of the hydrogen model for describing the excitons. With increasing the field excitonic states with different parity become mixed, leading to optical activation of states that are dark in zero field. As compared with atoms, due to the reduced Rydberg energy states with different $n$ can be brought into resonance in the accessible electric field strength range. When this occurs, we observe mostly crossing of levels within the experimental accuracy showing that the electron and hole motion remains regular. The observed features are well described by detailed calculations accounting for the spin-orbit coupling, the cubic anisotropy effects, and the symmetry-imposed optical selection rules.
\end{abstract}

\maketitle

\section{Introduction}\label{sec:intro}

 Application of electric and magnetic fields has turned out to be extremely helpful for developing a detailed understanding of atoms, as the electronic level structure is modified by characteristic energy shifts and/or lifting multiple state degeneracies. The latter is associated with the symmetry reduction from the SO(3) group of all rotations down to U(1) for rotations about the field, caused by introduction of the preferential field direction in space. Typically, the separation between atomic levels belonging to different principal quantum numbers $n$ is so large that resonances between these states can hardly occur for the field strengths accessible in the laboratory. This is different for Rydberg atoms where the valence electron has been promoted into a level with high $n \gg 1$~\cite{Gallagher}. In this range, for the hydrogen atom the levels of the Stark ladder simply cross each other, when they come into resonance by application of an electric field. By contrast, for many-electron atoms the excited electron experiences close to the nucleus a deviation from the $1/r$-Coulomb potential due to the core electrons. The influence of this deviation is often comprised in the quantum defect, by which the Rydberg formula for the excited electron binding energy is modified to ${\mathcal R} / \left( n - \delta_{n,l} \right)^2$, where ${\mathcal R}$ is the Rydberg energy and $\delta_{n,l}$ is the quantum defect for an electron with orbital angular momentum $l$. Accordingly, the states within a particular shell show a splitting in zero field, and when an electric field is applied they may show avoided crossings when they approach each other.

Excitons in semiconductors can be described as hydrogen-like complexes for which the crystal environment is captured by introducing effective carrier masses and the dielectric function, resulting in a renormalization of the Rydberg energy, $\mathcal R \to \mathcal R^*$. Like in atomic physics, electric and magnetic fields are essential tools for obtaining information about the details of the energy spectrum, determined also by the material parameters. However, typically the exciton Rydberg energy is quite small, so that the observable states are limited to small principal quantum numbers, for example, $n \lesssim 3$ in the prototypical direct band gap semiconductor GaAs. Moreover, when applying an external field it is challenging to separate the split states in energy because of their proximity, making a state identification complicated. In addition, exciton ionization occurs already in relatively weak electric fields.

Here, Cu$_2$O has a somewhat unique position because of its remarkable crystal quality in combination with a large exciton Rydberg energy of about $\mathcal R^* =$90 meV~\cite{gross:exciton:eng}. Both features have allowed us to observe exciton states up to $n$ = 25 in high resolution laser absorption studies~\cite{Kazimierczuk:2014yq}. These states, even in absence of  external fields, show a fine structure splitting that is related to the reduction from the continuous symmetry of atoms to the discrete O$_h$ symmetry of the cubic crystal lattice~\cite{PhysRevLett.115.027402}. The fine structure can be traced to the details of the band structure where band mixing effects lead to considerable deviations from a simple parabolic dispersion mostly for the holes~\cite{PhysRevLett.115.027402,PhysRevB.93.075203,PhysRevB.93.195203}. By analogy with atoms, the resulting exciton level series can be approximately cast into an effective quantum defect model, even though we stress that the origin of the quantum defect is quite different~\cite{PhysRevB.93.075203}.

Despite the sizable Rydberg energy compared to other semiconductors, the splitting between different exciton states with $n\geqslant 3$ is small enough, that application of feasible electric fields should allow one to bring states with different $n$ close to each other or even in resonance, without going to highly excited Rydberg states. From the quantum defect analogy a considerable state mixing and avoided crossings may occur between levels. Naively one may still expect that the absorption spectrum has the rather simple appearance of a Stark ladder, as the optical parameters like the dielectric function are scalar quantities for a cubic crystal. Here we present the results of an extensive experimental study of the excitonic absorption spectrum for the states with low principal quantum numbers $n$ = 3, 4, and 5 in Cu$_2$O crystals in presence of an electric field applied along the light propagation axis. By contrast to the simple expectations, we observe a rich fine structure of the spectra with intricate dependences on crystal orientation and light polarization. This is because the electric field reduces the system symmetry further down to at least $C_{4v}$ and thereby shuffles oscillator strength from the dominant $P$-excitons with $l=1$ to other angular momentum states, activating dipole-forbidden transitions~\cite{PSSA:PSSA2210430102}. Additionally, the involvement of quadrupolar transitions leads to a high complexity of the spectra involving multiple lines whose presence in the spectra depends strongly on the polarization of the incident laser light relative to the crystal orientation. Our experimental findings are substantiated by detailed numerical calculations.

For completeness, we note that the cuprous oxide exciton level spectrum in electric field has been experimentally studied before, see the original papers~\cite{GrossZakharchenya,PSSB:PSSB2220660140} and reviews \cite{PSSA:PSSA2210430102,Gross:1956aa}. However, in these studies the fine structure could not be resolved in that much detail, which may be related, for example, to the application of much higher fields 
in order to observe similar effects. Also a connection to previous microscopic calculations~\cite{PSSB:PSSB2220660140,Ziemkiewicz} could not be established. Here the understanding of the exciton level structure is taken to an unprecedented level by providing studies with high spectral resolution and comparison with model calculations which take into account, besides Coulomb potential and static electric field, symmetry allowed effects, such as the spin-orbit coupling, the cubic anisotropy effects and the short-range electron hole exchange interaction.

The paper is organized as follows: In the next section we give the details of the experiment. In Section~\ref{sec:results} we present the experimental results, while in Section~\ref{sec:theory} the theoretical model for the excitons in electric field is described. The comparison between theory and experiment is provided in Section~\ref{sec:disc}.


\section{Samples and experimental setup}\label{sec:samples}

In our absorption studies natural cuprous oxide crystal slabs of different orientations were studied: one sample (sample 1 with a thickness of 50 $\mu$m) was cut and polished such that the [001] direction is along the crystal normal; in the other case (sample 2 with a thickness of 30 $\mu$m) the [110] direction is normal to the surface. The samples were inserted in the liquid Helium insert of an optical cryostat at $T$ = 1.3 K.

The optical axis was chosen along the crystal normal. Excitation was done by using a broadband white light source the bandwidth of which was reduced to the wavelength range of interest by a double monochromator.
The polarization of the exciting light was chosen to be linear, so that for propagation of the beam along the [001] direction in sample~1, the polarization vector $\hat{\bm e}$ was along [100], equivalent to $\hat{\bm e}\parallel$[010]. For the sample~2 with the light propagating along the [110] direction, the two different independent light polarizations are $[001]$ and $[1{\bar1}0]$. It is noteworthy that the $[001]$ and $[1\bar 10]$ axes are not equivalent in the cubic $O_h$ point group, hence, it is not a priori obvious whether the transmission spectra are identical or not for these two polarization configurations.

After transmission through the crystal the light is dispersed by a double monochromator and detected by a Si-charge coupled device camera. The spectral resolution provided is less than 10~$\mu$eV. As will be explained in further detail below, we were focussing in our experiments on the excitons with
principal quantum number from $n=3$ to 5. From previous experiments~\cite{Kazimierczuk:2014yq,PhysRevLett.115.027402,Aszmann:2016aa} we know that the spectral widths of these features exceed considerably the above spectral resolution. Therefore we decided that the white light absorption is sufficient to study the electric field induced modification of the exciton spectrum. Indeed, up to a principal quantum number of $n =7$ the spectra for laser and white light excitation coincide. For comparison and in depth studies we have also used a tunable single frequency dye laser with a photo-diode detection system, see Sec.~\ref{sec:results} for details.

\begin{figure}
\includegraphics[width=0.8\linewidth]{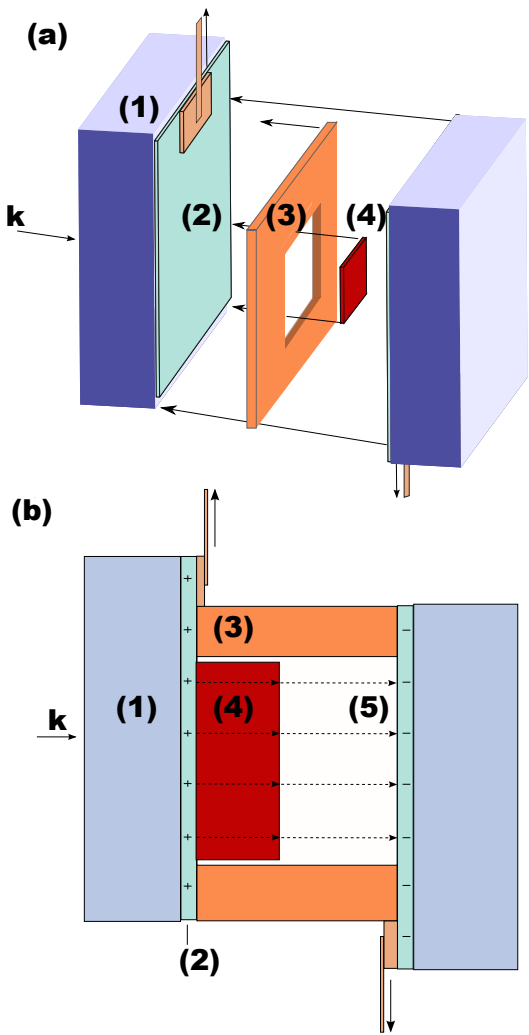} 
\caption{(Color online). Schematic sketch of the sample holder in exploded view (a) and view from the side (b): quartz-plates (\textbf{1}),  ITO-layers (\textbf{2}), kapton-spacer of 75~$\mu$m (\textbf{3}),  sample (\textbf{4}), free space filled with helium (\textbf{5}).}
\label{fig:sampleholder}
\end{figure}

For application of the electric field $\bm E$ a special holder was fabricated, in which the samples were mounted strain-free. The holder was designed such that the field could be applied along the optical axis (longitudinal configuration, $\bm{k}||\bm{E}$). By contrast, in previous experiments~\cite{PSSA:PSSA2210430102,Gross:1956aa,PSSB:PSSB2220660140}, only the transversal configuration with $\bm E \perp \bm k$ was studied. The sample holder is shown in Fig.~\ref{fig:sampleholder}. The crystal is mounted freely within a kapton-spacer of 75 $\mu$m thickness that is placed between quartz plates with transparent electrodes of indium tin oxide (ITO) evaporated on them. The thickness of the electrodes was chosen to minimize interference signals in the transmission spectra. We carefully checked that no measurable leakage current is flowing between the contacts.

We note that for recent fine structure studies of the exciton states this geometry was used to activate the $S$-  and $D$-excitons ($l=0$ and $2$, respectively) and extrapolate their energies towards zero electric field. However, in these studies the level splitting by electric field was not assessed~\cite{PhysRevB.93.075203}.

Despite all these efforts, we note that we observe depolarization effects in the optical measurements when applying an electric field. These effects may arise from charges on the crystal surfaces, charge redistribution among defects or optically injected carriers. As time resolved measurements have shown, these effects can be minimized by cutting down the laser power to less than $1~\mu$W and pulsing the electric field. These tools therefore were consequently applied in our studies. With these low laser powers the effects of the electric field on the spectra were the same in laser excitation as well as in white light excitation. In particular, we emphasize that the excitation densities are so low, that the formation of dressed states by simultaneous driving of two closely spaced exciton levels by a laser as demonstrated recently for Rydberg excitons, see Ref.~\cite{PhysRevLett.117.133003}, is not relevant.

Nevertheless, we find a difference in the electric field scale in the calculations and the scale in experiment determined as described above. Since the precise influence of these depolarization effects is hard to determine, we have rescaled the fields used in the calculations by a constant (field-independent) factor when comparing them with experiment. The scaling factors are about 3 \ldots 5 and do not vary systematically with studied sample, exciton principal quantum number etc. Most likely the variation of the scaling factors is related to the depolarization-causing charge distributions, which may vary upon thermal cycling of the crystal, for example.
Still, we stress again that the elaborated design of the sample holder and the careful choice of experimental conditions have allowed the observation of field-induced effects at field strengths much smaller than in earlier studies.

\section{Experimental results}\label{sec:results}

\subsection{Phenomenological analysis}

We study the exciton states of the yellow exciton series in cuprous oxide, involving the topmost valence band and the lowest conduction band. These bands are formed mainly from copper orbitals: The conduction band is of $s$-type character and the valence band is of $d$-type character, both with even parity. Optical transitions in electric-dipole approximation require an orbital angular momentum component change by unity, $\Delta l = \pm 1$, which obviously cannot be achieved in band-to-band transitions at the $\Gamma$ point, i.e., the center of the Brillouin zone. The total angular momentum and resulting symmetry of the exciton are given by the product of electron and hole Bloch amplitudes with the envelope function. Hence, in contrast to other semiconductors such as GaAs or CdTe, the dipole-transitions in cuprous oxide become possible due to a nonzero angular momentum of the envelope, namely by exciting excitons with a $P$-type envelope wavefunction~\cite{PhysRevLett.115.027402,PSSB:PSSB2220660140}.

The spin-orbit interaction results in the complex coupling of the envelope and Bloch functions angular momenta~\cite{PhysRevLett.115.027402,PhysRevB.23.2731}. Hence, one expects the exciton with a certain principal quantum number $n$ to consist of a multiplet comprising a large number of states. The symmetry reduction from SO(3) to $O_h$ by the cubic crystal environment, as compared to the isotropic model often used to describe excitons from the atom analogy, has been shown to activate higher exciton states with odd parity of the envelope: In a crystal angular momentum is strictly speaking no longer a good quantum number. As a result, oscillator strength is shuffled from the dominant $P$-excitons to the $F$- or $H$-excitons which therefore become detectable in single photon absorption as demonstrated recently~\cite{PhysRevLett.115.027402}. Still, the oscillator strength of these features is comparatively small. Therefore we will use the angular momentum notation for the excitons at zero field, for reasons of simplicity and to facilitate the connection to previous results.

The effect of a static electric field is described by the Hamiltonian~\cite{PSSB:PSSB2220660140,Ziemkiewicz}
\begin{equation}
\label{E:eff}
{\mathcal H_E}=-e \bm{r} \cdot \bm{E},
\end{equation}
acting on the exciton envelope functions, where $\bm r$ is the coordinate of the relative electron-hole motion. The electric field has odd parity with respect to the space inversion. As a result, the electric field lifts the inversion symmetry of the cuprous oxide crystal. If applied along the high symmetry crystallographic direction $[001]$, the symmetry is reduced to the $C_{4v}$ point symmetry group. This will lead to an additional state mixing, by which the odd envelope $P$-excitons are mixed with the even $S$- and $D$-excitons in lowest order perturbation theory. Similarly, the $F$-excitons become mixed with the even envelope $G$-excitons (as well as with $D$-excitons), provided the principal quantum number permits such high angular momentum states according to the limitation to $l_{\rm max}= n-1$. As a consequence of this state mixing, by application of the electric field not only a splitting of the lines observed already at zero field occurs, but also multiple additional lines are expected to emerge, resulting in a Stark ladder. If applied along the [110] direction (as in the sample 2), the symmetry is even further reduced to $C_{2v}$ point symmetry from which we expect a further enhancement of the effects of state mixing.

\subsection{Results of experiments}

\begin{figure}%
\includegraphics[width=0.99\linewidth]{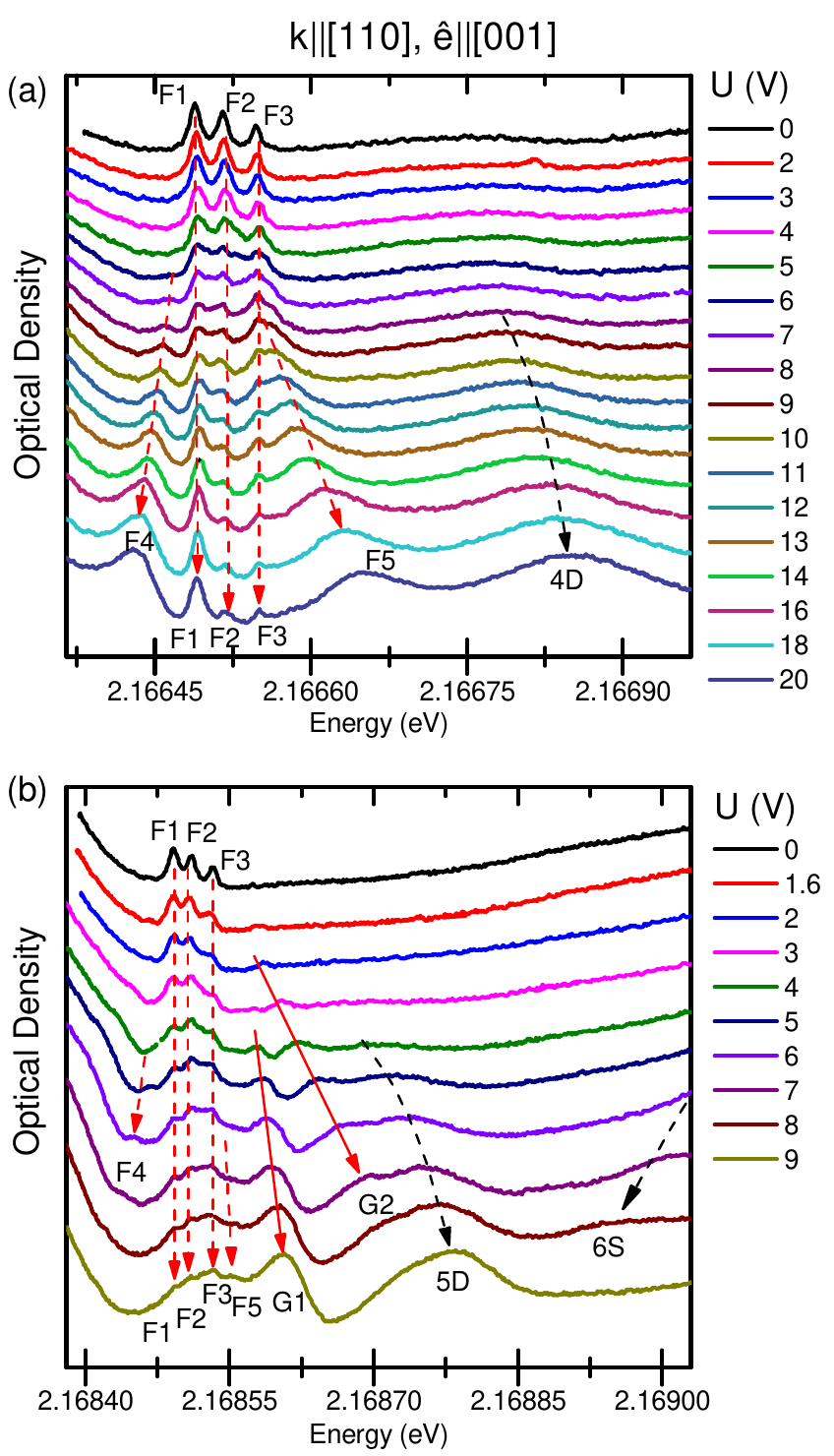}%
\caption{(Color online). High resolution absorption spectra of the high energy side of (a) the $4P$-resonance and (b) the $5P$-resonance for different applied voltages, recorded on sample 2 with light polarized along $[001]$. For $n=4$ the $F$-excitons split into $5$ lines. For $n=5$ we observe two additional lines besides the $F$-states, which are identified as $G$-envelope excitons. Note the different scales at the energy axis. The increase of optical density towards lower energies corresponds to the high energy flanks of the $P$-excitons.}%
\label{fig:FG-exc}%
\end{figure}

\begin{figure*}%
\includegraphics[width=0.75\linewidth]{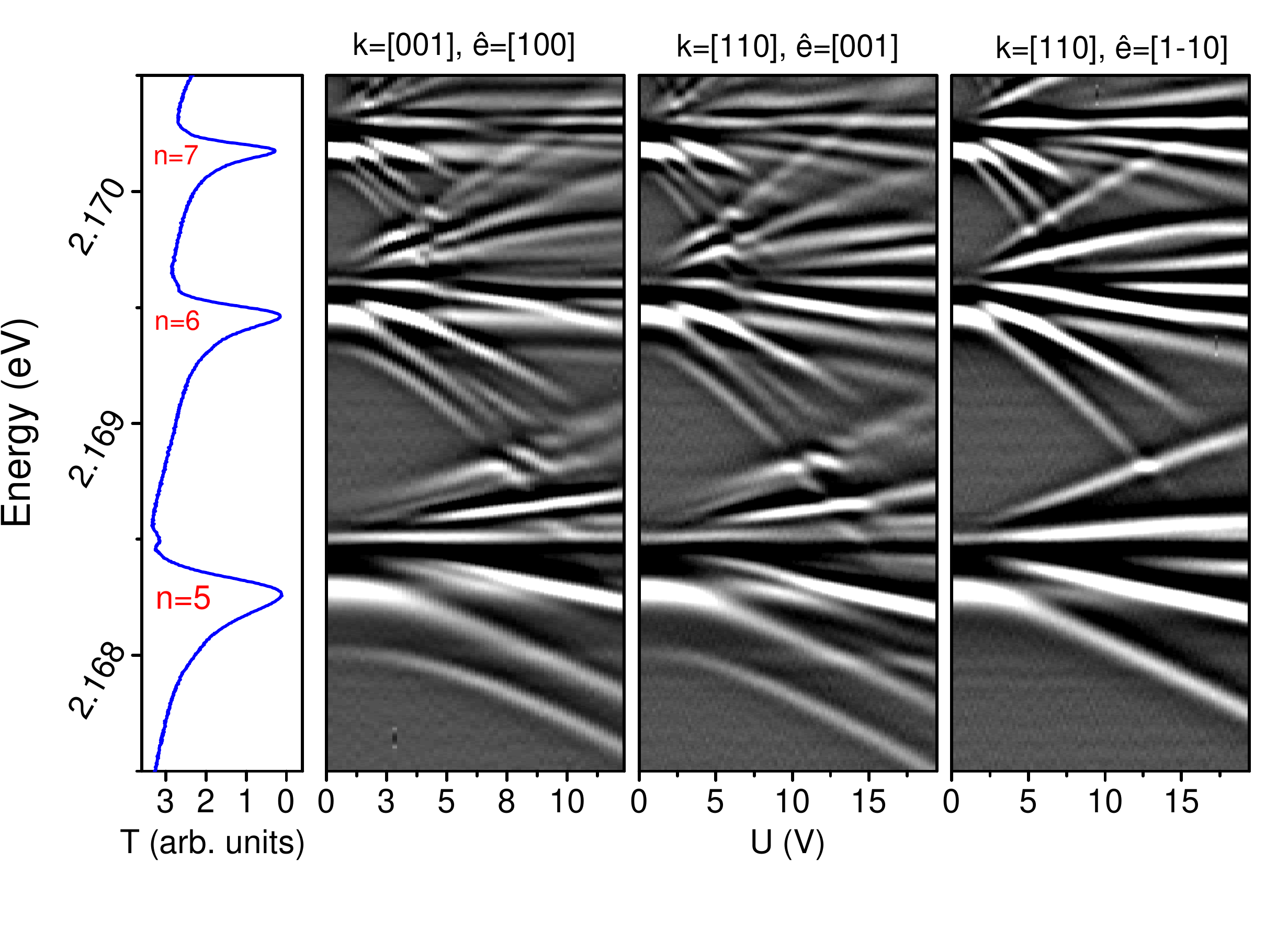}%
\caption{(Color online). Contour-plots of the second derivative of transmission spectra covering the energy range from $n=5$ to $n=7$ as function of applied voltage. For reference the leftmost panel shows the zero field spectrum which is almost entirely dominated by the $P$-excitons. The contour plots are shown for sample 1 with [001] orientation and light polarized along [100], and for sample 2 with orientation [110] for light polarizations [001] and $[1\bar10]$, respectively. 
}%
\label{fig:comparison}%
\end{figure*}

To test these expectations and to set a reference for the subsequent white light studies, we recorded electric field series of the high energy side of the $n$ = 4 and 5 excitons using a tunable single frequency dye laser (Sirah Matisse DS) with a linewidth of about 5~neV \cite{Kazimierczuk:2014yq}. The laser emission was stabilized by a noise eater (TEM Messtechnik NoiseEater) in front of the sample with a suppression of intensity fluctuations by about a factor $200$. The transmitted laser light was detected by a photo-diode. The corresponding spectra are shown in Fig.~\ref{fig:FG-exc} as water-fall plots. Since, in general, the spectra are dominated by the $P$-excitons with an intensity at least two orders of magnitude larger than that of any other line (see also the left hand side of Fig.~\ref{fig:comparison}, giving the transmission for the excitons from $n$ = 5 to 7), we have cut the $P$-excitons from the shown energy ranges and present only their high energy flanks for different applied voltages, in order to focus on the weaker lines from excitons with higher angular momenta.

The triplet of $F$-excitons activated by the mixing with the $P$-excitons due to the spin-orbit interaction and cubic symmetry effects are clearly observable  
starting from $n$=4 onwards. Previous calculations showed that four $F$-excitons should have finite oscillator strength, but two of them are almost degenerate~\cite{PhysRevLett.115.027402,PhysRevB.93.195203}. Application of the field lifts this degeneracy and leads also to activation of other states. In the $n$=4 spectra we observe two more $F$-exciton states appearing whose character we conclude from the emergence in the same energy range as the triplet. Simultaneously one of the original $F$-exciton peaks remains prominent, while the other two become faint. Note that the linewidths of the additional exciton resonances are considerably broader than those of the original ones, and are similar to those of the $P$-excitons. On the high energy side also another feature appears which we tentatively attribute to a $D$-exciton state, also with quite large linewidth.

Similar absorption lines with qualitatively the same appearance can be resolved for the multiplet related to the $n$ = 5 exciton. Note, however, that as expected from the larger polarizability of this exciton the equivalent splitting patterns appear at considerably smaller fields, roughly reduced by a factor of 2 illustrating strongly superlinear scaling of the polarizability  with the principal quantum number~\cite{ll3_eng}.
Also additional lines appear compared to the $n$=4 case. As increase of the principal quantum number by unity also allows for a larger value of the maximum angular momentum $l_{\rm max}$ by unity, we attribute these two lines to $G$-excitons with angular momentum $l$=4.
Note that for $n$ = 5 also a broad spectral feature shifting from higher to lower energies with increasing voltage appears on the high energy flank. We attribute this line to the $S$-exciton from the $n = 6$ multiplet.

Irrespective of these details, the two electric field series confirm the expectations of a complex level splitting pattern which cannot be explained within a simplified hydrogen-like model.

To obtain better insight into the fine details of the spectra with all the faint lines and the strong field-induced changes, we have calculated the second derivative over the energy of every recorded spectrum and assembled them in a contour plot. Doing so, the huge differences in intensity between the different features are levelled and an easier comparison becomes possible. This is shown in Fig. \ref{fig:comparison} for the excitons with principal quantum numbers from $n=5$ to $n=7$ for three different cases, sample 1 with [001] orientation and [100] light polarization (left contour plot) and sample 2 with [110] orientation, in one case for [001] polarization (mid) and in the other case for $[1\bar10]$ polarization (right), respectively. Here the studied energy range has been on purpose extended also to excitons with $n > 5$, in order to present a systematic phenomenology. For the quantitative comparison with theory we will later on restrict to $n = 3$ to $5$, as for higher principal quantum numbers accurate calculations are challenging due to the huge size of the Hilbert space involving multiple states, in particular, because of the vicinity of the states with different $n$ and necessity to account for ionization processes, which we did not include in our modeling.

Comparing the contour plots in Fig.~\ref{fig:comparison}, one finds the somewhat surprising result: In the cubic crystal of cuprous oxide the transmission spectra depend on the light propagation direction and light polarization. The number of lines vary from panel to panel in Fig.~\ref{fig:comparison} and the spectra have the simplest appearance for $\bm k \parallel [110]$ and $\hat{\bm e} \parallel [1\bar 10]$ coming closest to the simple expectations of a Stark ladder. It is even more surprising that this particular light propagation direction corresponds to the lowest symmetry direction described by the $C_{2v}$ point group.  On the other hand, the spectra for the other two configurations from first sight appear rather similar, but deviate considerably from a Stark fan.

In addition, we note that within the experimentally achievable fields exciton levels from multiplets of different principal quantum numbers can be brought into resonance, especially for the higher $n$. In the right panel, within the experimental accuracy the levels cross each other, contributing to the rather simple appearance of the contour plot.
This crossing behavior changes for the configurations in the other two contour plots: Here, in particular, on the high energy flank of a $P$-exciton anticrossings are observed, when the states with given $n$ (moving to higher energies) come into resonance with the states descending in energy from the exciton with the next
higher principal quantum number. As mentioned above, the anticrossings 
are somewhat expected from the reduced symmetry in the crystal, effects of the spin-orbit interaction and the deviation of the Coulomb interaction potential in the exciton from a pure $1/r$-form by the presence of the short-range part of the electron-hole interaction.
One also confirms that the field induced splitting between different states within a multiplet with a given $n$ becomes visible at the lower fields, the higher the principal quantum number is. In addition, the exciton lines split in more and more states with increasing $n$ as higher angular momenta become accessible.

\section{Theory}\label{sec:theory}

In order to describe the experimental data and explain the dependence of the observed transmission spectra on the light propagation direction and the experimental geometry, in this section we present the model of the exciton Stark effect in Cu$_2$O and address the optical selection rules for excitonic transitions in the presence of the electric field.

\subsection{Effective Hamiltonian}\label{sec:Heff}

We consider the effect of the external electric field $\bm E$ on the yellow exciton series formed of an electron with $\Gamma_6^+$ symmetry Bloch amplitude and a valence band hole with $\Gamma_7^+$ Bloch amplitude. 
The complex valence band structure results in the admixture of the $\Gamma_8^+$ valence band to the $\Gamma_7^+$ band in the exciton~\cite{PhysRevB.23.2731,PhysRevLett.115.027402,PhysRevB.93.195203}.
The admixture yields (i) the fine structure splitting of exciton multiplets with a given principle quantum number $n$ and (ii) the fine structure of states with a given orbital angular momentum $l$~\cite{PhysRevLett.115.027402}. In accordance with Eq.~\eqref{E:eff} the main effect of the external electric field is on the exciton envelope functions, hence, the band mixing effects can be included in the hydrogen-like model perturbatively~\cite{Ziemkiewicz}. Therefore, we introduce the basis of the exciton states
\begin{multline}
\label{basis}
|n,l,m,s_e,s_h\rangle  = |n,l,m\rangle |s_e,s_h\rangle \\
= R_{nl}(r) Y_{lm}(\vartheta,\varphi) |s_e,s_h\rangle,
\end{multline}
where $\bm r$ is the position-vector of the relative electron-hole motion, $r=|\bm r|$, $\vartheta$ and $\varphi$ are its polar and azimuthal angles, respectively, $R_{nl}(r)$ are the hydrogen-like radial functions, and $Y_{lm}(\vartheta,\varphi)$ are the spherical harmonics~\cite{ll3_eng}. The factor $|s_e,s_h\rangle$ is the product of the electron and hole Bloch amplitudes with $s_e,s_h=\pm 1/2$, being the spin components.
%
%
The effective exciton Hamiltonian in cuprous oxide can be written as
\begin{equation}
\label{H}
\mathcal H = \mathcal H_0 + \mathcal H_{so} + \mathcal H_{exch} + \mathcal H_{cubic} + \mathcal H_E,
\end{equation}
where $\mathcal H_0$ takes into account only the orbital energies, $\mathcal H_{so}$ is the spin-orbit coupling Hamiltonian, $\mathcal H_{exch}$ describes the exchange interaction between the electron and the hole and $\mathcal H_{cubic}$ comprises the effects of the cubic ({$O_h$} point symmetry) environment. In our calculations, the excitonic wave function is decomposed over the basis functions~\eqref{basis}, and the Hamiltonian $\mathcal H$, Eq.~\eqref{H}, is represented in matrix form. The diagonalization of this matrix yields the exciton energies and wave functions which enable us to calculate the transmission spectra in the electric field. 

The orbital Hamiltonian is diagonal in the quantum numbers $n$, $l$, $m$, $s_e$ and $s_h$ and its non-zero matrix elements read
\begin{equation}
\label{orbital}
\langle n,l,m, s_e, s_h|\mathcal H_0 |n,l,m, s_e, s_h\rangle = E^0_{nl},
\end{equation}
where the diagonal energies $E^0_{nl}$ can be conveniently parametrized in terms of the exciton Rydberg and the quantum defect:
\begin{equation}
\label{diagonal}
E^{0}_{nl} = \frac{\mathcal R^*}{(n-\delta_{nl})^2}.
\end{equation}
The spin-orbit coupling is dominant for the valence band states. Therefore it is sufficient to take into account the interaction of the hole spin and its orbital momentum only. In the spherical approximation the spin-orbit Hamiltonian acting in the manifold of the states with given $n$ and $l$ reads
\begin{equation}
\label{H:so}
\Delta^{(so)}_{nl} (\hat{\bm s}_h \cdot \hat{\bm l}),
\end{equation}
where $\Delta^{(so)}_{nl}$ are the corresponding spin-orbit coupling constants, $\hat{\bm s}_h=\bm \sigma_h/2$ is the hole spin operator with $\bm \sigma_h = (\sigma_{h,x},\sigma_{h,y},\sigma_{h,z})$ being the Pauli matrices acting on the pair of hole Bloch amplitudes $|\Gamma_7^+,\pm 1/2\rangle$ and $\hat{\bm l}=(l_x,l_y,l_z)$ is the orbital angular momentum operator acting in the space of spherical harmonics, e.g., $\hat l_z Y_{lm} = mY_{lm}$.

The effective Hamiltonian $\mathcal H_1=\mathcal H_0 + \mathcal H_{so}$ is quite similar to that of Rydberg atoms~\cite{PhysRevA.20.2251}. It has spherical symmetry and allows for classification of the excitonic states by the atom-like nomenclature: the principle quantum number $n$, the angular momentum $l$, and the total momentum $f$, where $\hat{\bm f} =  \hat{\bm s}_h + \hat{\bm l}$, its $z$-component $f_z$ and the electron spin component $s_e$. The eigenstates of $\mathcal H_1$ are degenerate in $f_z$ and $s_e$, their wave functions and energies read~\cite{ll3_eng}
\begin{subequations}
\label{states:H1}
\begin{equation}
\label{psi:H1}
|n,l,f,f_z,s_e\rangle = \sum_{m,s_h} C^{ff_z}_{lm;1/2s_h} |n,l,m,s_e,s_h\rangle,
\end{equation}
\begin{equation}
\label{E:H1}
E^1_{nlf}=E_{nl}^0 + \frac{\Delta^{(so)}_{nl}}{2}\left[f(f+1)-l(l+1)-3/4\right].
\end{equation}
\end{subequations}
Here the $C^{ff_z}_{lm;1/2s_h}$ are the Clebsch-Gordan coefficients. As a result, neglecting $\mathcal H_{exch}$ and $\mathcal H_{cubic}$ one can denote the excitonic states as, e.g., $1S_{1/2}$, $2P_{1/2}$, $2P_{3/2}$, $3D_{3/2}$, $3D_{5/2}$, etc., by the principle quantum number $n$, the symmetry of the envelope function ($S$ for $l=0$, $P$ for $l=1$, etc.) and the total angular momentum $f=l\pm 1/2$. Microscopically, the spin-orbit interaction results from the complex valence band structure, and  particularly, from the mixing of the $\Gamma_7^+$ and $\Gamma_8^+$ valence bands~\cite{PhysRevB.23.2731,PhysRevLett.115.027402}.

By contrast, the contributions $\mathcal H_{exch}$ and $\mathcal H_{cubic}$ are specific for excitons in semiconductors. $\mathcal H_{exch}$ describes the exchange interaction between the electron and the hole forming the exciton. It can be decomposed in the long- and short-range parts~\cite{BP_exch71}. The former is relatively weak for ``forbidden'' transitions and is disregarded here, see Refs.~\cite{PhysRevLett.91.107401,PhysRevB.93.195203,PhysRevB.94.115201} for details. In what follows we consider only the short-range part of the exchange interaction, $\propto \delta(\bm r)$, where $\delta(\bm r)$ is the three-dimensional $\delta$-function. This interaction is relevant for the $S$- and $D$-shell excitons: for the $S$-shell states the envelopes $R_{n0}(r)$ do not vanish at $r\to 0$, and the $D$-shell states are strongly mixed with the $S$-shell states from the $\Gamma_8^+$ valence band, see Ref.~\cite{PhysRevB.23.2731} for details. Within the $S$- and $D$-multiplets the exchange interaction is described by the constants $J^s_n$ and $J^d_n$, respectively: 
\begin{equation}
\label{exch:ss:dd}
J^S_n(\hat{\bm s}_e \cdot \hat{ \bm s}_h) ,  \quad J^D_{nf}(\hat{\bm s}_e \cdot \hat {\bm f}).
\end{equation}
With account for the contributions in Eq.~\eqref{exch:ss:dd} the exciton states are split in accordance to the total angular momentum of the electron and the hole {$ \hat{\bm J} =  \hat{\bm s}_e +  \hat{\bm f}$}. In particular, the $S$-shell states split into the paraexciton $J=0$ and the triplet orthoexciton $J=1$, while the $D_{3/2}$-states split into states with $J=1,2$ and the $D_{5/2}$-states split into $J=2$ and $J=3$ multiplets. We note, however, that the exchange interaction is most important for the $D_{3/2}$-states, while for the $D_{5/2}$ as well as for the $F$-, $G$-, etc. excitons, the effects of the symmetry reduction due the cubic crystalline environment are important, see Ref.~\cite{PhysRevLett.115.027402} and below for details.  Moreover, the exchange interaction mixes the $S$- and $D$-shell states with $J=1$, as well as the {$D$}-shell multiplets with $J=2$. For the following the mixing of the $J=2$ states is disregarded and the mixing of the $S$-shell and $D$-shell states with $J=1$ is characterized by the parameters
\begin{equation}
\label{exch:sd}
J^{SD}_n=\langle nS,J=1,J_z|\mathcal H_{exch}|nD_{3/2},J=1,J_z\rangle,
\end{equation}
with $J_z$ being the $z$-component of the total angular momentum $\bm J$.

The $\mathcal H_{cubic}$ term reduces the symmetry of the problem from spherical to the {$O_h$} point group, resulting in mixing and splitting of the states with given total angular momentum. 
We illustrate the role of these terms by the example of the $D_{5/2}$ excitonic states. In the $O_h$ point symmetry group these states transform according to the reducible representation $\Gamma_6^+ \oplus \Gamma_8^+$. The crystalline splitting between these states is described by the effective Hamiltonian
\begin{equation}
\label{cubic}
\mathcal H_{cubic,nd} =  \frac{\delta_{cubic,nd}}{18} (\hat f_x^4+ \hat f_y^4+ \hat f_z^4),
\end{equation}
where $\delta_{cubic,nd}$ is the splitting between the $\Gamma_8^+$ quadruplet and the $\Gamma_6^+$ doublet, $\hat{\bm f}$ are the matrices of the angular momentum $5/2$ in the set of cubic axes. Similar invariants can be written for the $F$-, $G$-, etc. excitons. Microscopically, the origin of this splitting is the difference of the $\gamma_2$ and $\gamma_3$ Luttinger parameters~\cite{PhysRevLett.115.027402,PhysRevB.93.195203,cubic}.

Finally, the matrix elements of the electric field-induced perturbation, Eq.~\eqref{E:eff}, are easily calculated making use of the general expressions for the matrix elements of the vector components in the basis of the $Y_{lm}(\vartheta,\varphi)$~\cite{Ziemkiewicz,edmonds} and the integrals of the radial hydrogenic wave functions~\cite{BS}. In our calculations we keep the momentum quantization axis parallel to the electric field ($z||[001]$ for comparing with the experimental data on sample 1 and $z'\parallel [110]$ for comparing with the data on sample 2). 

\subsection{Selection rules for light propagating along the cubic axis}\label{sec:selection:001}

In order to calculate the transmission spectra one needs to determine the exciton oscillator strengths. We take into account both dipole transitions to $P$-shell excitonic states and quadrupolar transitions to $S$-shell states. To that end we introduce the three orbital Bloch amplitudes of the valence band $\mathcal U_\alpha$, $\alpha=x,y,z$ which transform, respectively, as $yz$, $xz$ and $xy$ in the $O_h$ point group, i.e. according to the $\Gamma_5^+$ irreducible representation. The Bloch amplitudes of the $\Gamma_7^+$ states in the electron representation can be written as
\begin{subequations}
\label{G7}
\begin{align}
|+1/2,\Gamma_7^+\rangle = -\sqrt{\frac{2}{3}} \frac{\mathcal U_x + \mathrm i \mathcal U_y}{\sqrt{2}}\downarrow - \frac{1}{\sqrt{3}} \mathcal U_z \uparrow, \\
|-1/2,\Gamma_7^+\rangle = -\sqrt{\frac{2}{3}} \frac{\mathcal U_x - \mathrm i \mathcal U_y}{\sqrt{2}}\uparrow + \frac{1}{\sqrt{3}} \mathcal U_z \downarrow.
\end{align}
\end{subequations}
Here $\uparrow$, $\downarrow$ are the spin-up and spin-down electron spinors (i.e., the basis functions of the $\Gamma_6^+$ irreducible representation) and the set of cubic axes is used with $x\parallel [100]$, $y\parallel [010]$ and $z\parallel [001]$. The conduction band states are simply expressed via the spherically-symmetric Bloch amplitude, $\mathcal S$, and spinors such as $|+1/2,\Gamma_6^+\rangle = S\uparrow$, $|-1/2,\Gamma_6^+\rangle = S\downarrow$. It is convenient to parametrize the matrix elements of the optical transitions in terms of basis orbital matrix elements involving one Bloch amplitude $ \mathcal U_\alpha$ and one orbital envelope of the exciton. Particularly, for the $P$-shell excitonic states (envelope $\Gamma_4^-$, $l=1$) the matrix elements for the interband dipole transition can be recast as~\cite{knox1963theory,PSSB:PSSB2220660140,PhysRevB.93.195203}
\begin{equation}
\label{dipole}
\langle \mathcal S|\hat{d}_\gamma|\mathcal U_\alpha \Phi_{n,\beta}\rangle =  \varkappa_{\alpha\beta\gamma} \mathcal D \left(\frac{R_{n1}(r)}{r} \right)_{r\to 0},
\end{equation}
where $\hat{\bm d}$ is the dipole operator, $\mathcal D$ is a material parameter~\cite{remote:kp,PhysRev.188.1294}, $\alpha$, $\beta$, $\gamma$ are the Cartesian components, $\Phi_{n,\beta}$ are the $P$-shell exciton envelopes transforming as the coordinate $\beta$ being a linear combination of functions~\eqref{basis} with fixed $n$ and $l=1$, and the symbol $\varkappa_{\alpha\beta\gamma}$ equals to $1$ if and only if all subscripts are different (otherwise it is zero). For light propagating along the $z$-axis the $s$-shell excitons are quadrupole-allowed in the $x$ and $y$ polarizations, and the matrix element of the transition reads
\begin{equation}
\label{quadrupole}
\langle \mathcal S|\mathrm i z\hat{d}_\gamma|\mathcal U_\alpha \Phi_{n,s}\rangle =  \varkappa_{\alpha z\gamma}\mathrm i \mathcal QR_{n0}(0),
\end{equation}
where $\Phi_{n,s}( r) = R_{n0}(r)/\sqrt{4\pi}$ is the $s$-shell envelope, and $\mathcal Q$ is another parameter~\cite{remote:kp,PhysRev.188.1294}. 

   \begin{table}[t]
\caption{Nonzero matrix elements of the quadrupolar transitions for light propagating along the cubic axis, $\bm k\parallel [001]$.}\label{tab:Q}
     \begin{ruledtabular}
     \begin{tabular}{c|c|c}
 $|exc\rangle$ & polarization & $\langle exc|\mathrm i  z \hat{d}_\gamma |0\rangle$ \\
 &  $\gamma$ & units of $\mathcal QR_{n0}(0)$\\
     \hline
 $nS_{1/2,-1/2};-1/2\Gamma_6^+$ & $x\parallel [100]$ & $ 1\over\sqrt{3}$\\
  $nS_{1/2,+1/2};+1/2\Gamma_6^+$ & $x\parallel [100]$ & $ 1\over\sqrt{3}$\\
   $nS_{1/2,-1/2};-1/2\Gamma_6^+$ & $y\parallel [010]$ & $ -\mathrm i \over\sqrt{3}$\\
  $nS_{1/2,+1/2};+1/2\Gamma_6^+$ & $y\parallel [010]$ & $ \mathrm i\over\sqrt{3}$
     \end{tabular}
 \end{ruledtabular}
     \end{table}
 
    \begin{table}[t]
\caption{Nonzero matrix elements of the dipole transitions for light propagating along the cubic axis, $\bm k\parallel [001]$.}\label{tab:D}
     \begin{ruledtabular}
     \begin{tabular}{c|c|c}
 $|exc\rangle$ & polarization  & $\langle exc| \hat{d}_\gamma |0\rangle$ \\
 & $\gamma$ & units of $\mathcal D\left(\frac{R_{n1}(r)}{r} \right)_{r\to 0}$\\
     \hline 
 $n P_{3/2,-3/2}; +1/2\Gamma_6^+$ & $x\parallel [100]$ & $ \frac{\mathrm i  }{\sqrt{6}}$\\
  $n P_{3/2,+1/2}; +1/2\Gamma_6^+$ & $x\parallel [100]$ & $ \frac{\mathrm i  }{\sqrt{2}}$\\
   $n P_{3/2,-3/2}; +1/2\Gamma_6^+$ & $y\parallel [010]$ & $ \frac{1 }{\sqrt{6}}$\\ 
  $n P_{3/2,+1/2}; +1/2\Gamma_6^+$ & $y\parallel [010]$ & $ -\frac{1}{\sqrt{2}} $\\
   $n P_{3/2,+3/2}; -1/2\Gamma_6^+$ & $x\parallel [100]$ & $ \frac{\mathrm i  }{\sqrt{6}} $\\
  $n P_{3/2,-1/2}; -1/2\Gamma_6^+$ & $x\parallel [100]$ & $ \frac{\mathrm i  }{\sqrt{2}} $\\
   $n P_{3/2,+3/2}; -1/2\Gamma_6^+$ & $y\parallel [010]$ & $ -\frac{ 1}{\sqrt{6}} $\\
  $n P_{3/2,-1/2}; -1/2\Gamma_6^+$ & $y\parallel [010]$ & $ \frac{1}{\sqrt{2}} $\\
  $n P_{5/2,-3/2}; +1/2\Gamma_6^+$ & $x\parallel [100]$ & $\frac{2\mathrm i }{\sqrt{5}} $\\
 $ n P_{5/2,+1/2}; +1/2\Gamma_6^+$ & $x\parallel [100]$ & $\frac{\sqrt{2}\mathrm i  }{\sqrt{5}} $\\
$ n P_{5/2,-3/2}; +1/2\Gamma_6^+$ & $y\parallel [010]$ & $\frac{2 }{\sqrt{5}} $\\
$ n P_{5/2,+1/2}; +1/2\Gamma_6^+$ & $y\parallel [010]$ & $-\frac{\sqrt{2} }{\sqrt{5}} $\\
$ n P_{5/2,+3/2}; -1/2\Gamma_6^+$ & $x\parallel [100]$ & $-\frac{2\mathrm i  }{\sqrt{5}} $\\
$ n P_{5/2,-1/2}; -1/2\Gamma_6^+$ & $x\parallel [100]$ & $-\frac{\sqrt{2}\mathrm i  }{\sqrt{5}} $\\
$ n P_{5/2,+3/2}; -1/2\Gamma_6^+$ & $y\parallel [010]$ & $\frac{2 }{\sqrt{5}} $\\
$ n P_{5/2,-1/2}; -1/2\Gamma_6^+$ & $y\parallel [010]$ & $-\frac{\sqrt{2} }{\sqrt{5}} $
     \end{tabular}
 \end{ruledtabular}
     \end{table}
 
In order to transform Eqs.~\eqref{dipole}, \eqref{quadrupole} to the exciton basis we carry out the time reversal, which we define in agreement with Eq. (2.8.2) of Ref.~\cite{edmonds}:
\begin{equation}
\label{ku}
\hat{\mathcal K}|f,f_z\rangle = (-1)^{f+f_z} |f,-f_z\rangle.
\end{equation}
The valence band hole state  $|+1/2\Gamma_7^+\rangle_h$ passes to the electron state $-|-1/2\Gamma_7^+\rangle_e$, while the state $|-1/2\Gamma_7^+\rangle_h$ is transformed into the electron state $|+1/2\Gamma_7^+\rangle_e$. The matrix elements of the quadrupolar transitions to exciton states are summarized in Tab.~\ref{tab:Q}. Hereafter $|0\rangle$ denotes the ground state of the crystal. Note that other matrix elements with $s_h\ne s_e$ vanish. According to Eq.~\eqref{ku} the transformation rules for the $P_{f f_z}$-shell states read $|P_{3/2,+3/2}\rangle_h \to -|P_{3/2,-3/2}\rangle_e$,  $|P_{3/2,+1/2}\rangle_h \to |P_{3/2,-1/2}\rangle_e$, $|P_{3/2,-1/2}\rangle_h \to -|P_{3/2,+1/2}\rangle_e$, $|P_{3/2,-3/2}\rangle_h \to |P_{3/2,+3/2}\rangle_e$. The corresponding matrix elements of the dipole transitions are presented in Tab.~\ref{tab:D}. In addition to the transitions involving $P_{3/2}$-states we have also included the contributions from the $P_{5/2}$-states arising from the $\Gamma_8+$ valence band since the mixing with these states determines the oscillator strength of, e.g., the $F$-excitons. 
Other transitions which are not listed in the table as well as the transitions involving the $P_{1/2}$-states are forbidden. The Tabs.~\ref{tab:Q} and \ref{tab:D} demonstrate that the matrix elements of the exciton-light interaction are similar for the $x||[100]$- and $y||[010]$-polarizations, because the cubic axes are equivalent in the $O_h$ point symmetry group. Therefore, for light propagating along the cubic axis the transmission spectra are polarization independent.


\subsection{Selection rules for light propagating along the [110] axis}\label{sec:selection:110}

    \begin{table}[t]
\caption{Nonzero matrix elements of the dipole transitions for light with $\bm k\parallel [110]$.}\label{tab:D1}
     \begin{ruledtabular}
     \begin{tabular}{c|c|c}
 $|exc\rangle$ & polarization & $\langle exc| \hat{d}_\gamma |0\rangle$ \\
 & $\gamma$ & units of $\mathcal D\left(\frac{R_{n1}(r)}{r} \right)_{r\to 0}$\\
     \hline
$ nP_{3/2,3/2};+1/2\Gamma_6^+$ & $y'\parallel [1\bar10]$ & $\mathrm i    \sqrt{\frac{2}{3}}  $\\
$ nP_{3/2,-3/2};-1/2\Gamma_6^+$ & $y'\parallel [1\bar10]$ & $ -\mathrm i   \sqrt{\frac{2}{3}}  $ \\
$ nP_{5/2,3/2};+1/2\Gamma_6^+ $ & $y'\parallel [1\bar10]$ &  $\mathrm i \sqrt{\frac{1}{5}}  $\\
$ nP_{5/2,-5/2};+1/2\Gamma_6^+ $ & $y'\parallel [1\bar10]$ &  $ - \mathrm i     $\\
$ nP_{5/2,-3/2};-1/2\Gamma_6^+ $ & $y'\parallel [1\bar10]$ & $ \mathrm i   \sqrt{\frac{1}{5}}  $\\
$ nP_{5/2,5/2};-1/2\Gamma_6^+ $ & $y'\parallel [1\bar10]$ & $ - \mathrm i     $\\
$ nP_{3/2,3/2};+1/2\Gamma_6^+$ & $x'\parallel [001]$ & $ \frac{ 1}{\sqrt{6}}  $\\
$ nP_{3/2,-3/2};-1/2\Gamma_6^+$ & $x'\parallel [001]$ & $ \frac{ 1}{\sqrt{6}}   $\\
$ nP_{3/2,-1/2};+1/2\Gamma_6^+ $ & $x'\parallel [001]$ & $ \frac{1 }{\sqrt{2}}   $\\
$ nP_{3/2,1/2};-1/2\Gamma_6^+ $ & $x'\parallel [001]$ & $  \frac{ 1}{\sqrt{2}}   $\\
$ nP_{5/2,3/2};+1/2\Gamma_6^+ $ & $x'\parallel [001]$ & $ \frac{ 1}{2\sqrt{5}}  $\\
$ nP_{5/2,-3/2};-1/2\Gamma_6^+ $ & $x'\parallel [001]$ & $ \frac{1 }{2\sqrt{5}}   $\\
$ nP_{5/2,-1/2};+1/2\Gamma_6^+ $ & $x'\parallel [001]$ & $  \frac{3 }{\sqrt{10}}   $\\
$ nP_{5/2,1/2};-1/2\Gamma_6^+ $ & $x'\parallel [001]$ & $ - \frac{3 }{\sqrt{10}}   $\\
$ nP_{5/2,-5/2};+1/2\Gamma_6^+ $ & $x'\parallel [001]$ & $  \frac{1 }{2}   $\\
$ nP_{5/2,5/2};-1/2\Gamma_6^+ $ & $x'\parallel [001]$ & $  -\frac{1 }{2}   $
     \end{tabular}
 \end{ruledtabular}
     \end{table}

The situation is different for light propagating along the $z'\parallel [110]$ axis. Here the two axes in the plane perpendicular to the light wave vector, $x'\parallel [001]$, $y'\parallel [1\bar 10]$, are not equivalent. In the coordinate frame ($x',y',z'$) the Bloch amplitudes for the $\Gamma_7^+$ valence band have the same form as in Eq.~\eqref{G7} but with the rotated orbital and spinor basis functions:
\begin{subequations}
\label{G7'}
\begin{align}
|+1/2,\Gamma_7^+\rangle' = -\sqrt{\frac{2}{3}} \frac{\mathcal U_x' + \mathrm i \mathcal U_y'}{\sqrt{2}}\downarrow' - \frac{1}{\sqrt{3}} \mathcal U_z' \uparrow', \\
|-1/2,\Gamma_7^+\rangle' = -\sqrt{\frac{2}{3}} \frac{\mathcal U_x' - \mathrm i \mathcal U_y'}{\sqrt{2}}\uparrow' + \frac{1}{\sqrt{3}} \mathcal U_z' \downarrow',
\end{align}
\end{subequations}
where $\uparrow'$ ($\downarrow'$) are the spinors with the spin projections $\pm 1/2$ onto the $z'$ axis and the orbital Bloch amplitudes in the rotated axes frame are introduced as:
\begin{equation}
\label{U'}
\mathcal U_{x'} = \mathcal U_z, \quad \mathcal U_y' = \frac{\mathcal U_x - \mathcal U_y}{\sqrt{2}}, \quad \mathcal U_z' = \frac{\mathcal U_x + \mathcal U_y}{\sqrt{2}}.
\end{equation}
These orbital basis functions transform as $\mathcal U_x' \propto {(z'^2-y'^2)}/{2}$, $\mathcal U_y'\propto -x'y'$, and $\mathcal U_z' \propto x'z'$.

Making use of Eq.~\eqref{U'} the general expressions~\eqref{dipole}, \eqref{quadrupole} can be easily transformed to the $(x',y',z')$ coordinate frame. One can readily check that in the $y'\parallel [1\bar 10]$ polarization only $P$-excitons are active, while quadrupolar transitions are forbidden. By contrast, in $x'\parallel [001]$ polarization the quadrupolar transitions are allowed and, hence, both $P$- and $S$-shell states are active. Taking into account that the only non-zero quadrupolar matrix element is $ \langle \mathcal S |z' d_{x'}|\mathcal U_z' \Phi_{n,s}\rangle $ [cf. Eq.~\eqref{quadrupole}] we obtain
\begin{subequations}
\label{quad:exc}
\begin{align}
\langle ns_{-1/2};+1/2\Gamma_6^+ |\mathrm i  z' d_x'|0\rangle = \frac{\mathrm i \mathcal Q}{\sqrt{3}}R_{n0}(0), \\
 \langle ns_{+1/2};-1/2\Gamma_6^+ |\mathrm i  z' d_x'|0\rangle = \frac{\mathrm i \mathcal Q}{\sqrt{3}}R_{n0}(0).
\end{align}
\end{subequations}
The non-zero matrix elements of the optical transitions involving $P$-excitons are summarized in Tab.~\ref{tab:D1}. It is worth stressing that $nP_{1/2}$-states are forbidden as well. For $x'$ polarization the selection rules for $P_{3/2}$-excitons are similar to those presented above in Sec.~\ref{sec:selection:001} for $x$ polarization. Somewhat different selection rules for $P_{5/2}$-shell states hold because this multiplet transforms according to the reducible representation of the $O_h$ point group.

In bulk semiconductors the normal modes of the electromagnetic field are the exciton-polaritons, mixed waves of photons and excitons. The strength of polariton effects is controlled by the longitudinal-transverse splitting, $\Delta_{LT}$, which  describes the splitting between the transversal excitons states (coupled with the light) and the longitudinal ones (decoupled from the light). The oscillator strength of $2P$-excitons is on the order of $10^{-6}$~\cite{nikitine} and for the $1S$-excitons it is about three orders of magnitude smaller~\cite{froh:exch}. For the excited states the oscillator strengths are accordingly smaller. Correspondingly, the longitudinal-transverse splitting for the $2P$-exciton is about $\Delta_{LT} \approx 10$~$\mu$eV~\cite{Ziemkiewicz}. This gives the scale of the polariton shifts in the spectra for the $P$-excitons (while for the $S$- and $D$-excitons the polariton shifts are considerably smaller). Since $\Delta_{LT}$ is much smaller as compared with the linewidths of the relevant $P$-excitons, the polariton effects can be neglected. Moreover, for the same reasons the effects of the analytical exchange interaction between electron and hole in the exciton can be disregarded~\cite{Kavoulakis,PhysRevLett.91.107401,PhysRevB.94.115201}.

In order to provide a semi-quantitative comparison with experiment putting the main emphasis on the exciton line positions and relative oscillator strengths of the different states we calculate the absorption coefficient of the sample
 assuming that each exciton eigenstate $N_x$ of the Hamiltonian~\eqref{H} provides a Lorentzian line in absorption
\begin{equation}
\label{abs}
\mathcal A(E;\hat{\bm e})= \sum_{N_x} \frac{|M(N_x; \hat{\bm e})|^2}{(E- E_{N_x})^2+\Gamma_{N_x}^2},
\end{equation}  
where $M(N_x; \hat{\bm e})$ is the optical matrix element in the polarization $\hat{\bm e}$ calculated making use of  Tabs.~\eqref{tab:Q} --- \eqref{tab:D1} and Eq.~\eqref{quad:exc} (the absolute values of the parameters $\mathcal D$ and $\mathcal Q$ are fitted to obtain relative intensities of the calculated lines similar to those in experiment, see Appendix~\ref{sec:appendix} for parameter values), $E_{N_x}$ is the exciton energy, and $\Gamma_{{N_x}}$ is its damping. A consistent calculation of the transmission spectra requires analysis of the polariton modes and additional waves related with the space dispersion~\cite{excitons,Ziemkiewicz,note:refl}. We also note that for precise modelling of the spectra one has to account for the interference of excitonic transitions with the phonon-induced background~\cite{toy,sch:phon}. The model presented here is simplified but, as we demonstrate below, is sufficient to account for most of experimental observations.


\section{Discussion}\label{sec:disc}

\begin{figure*}[t]%
\includegraphics[width=\textwidth]{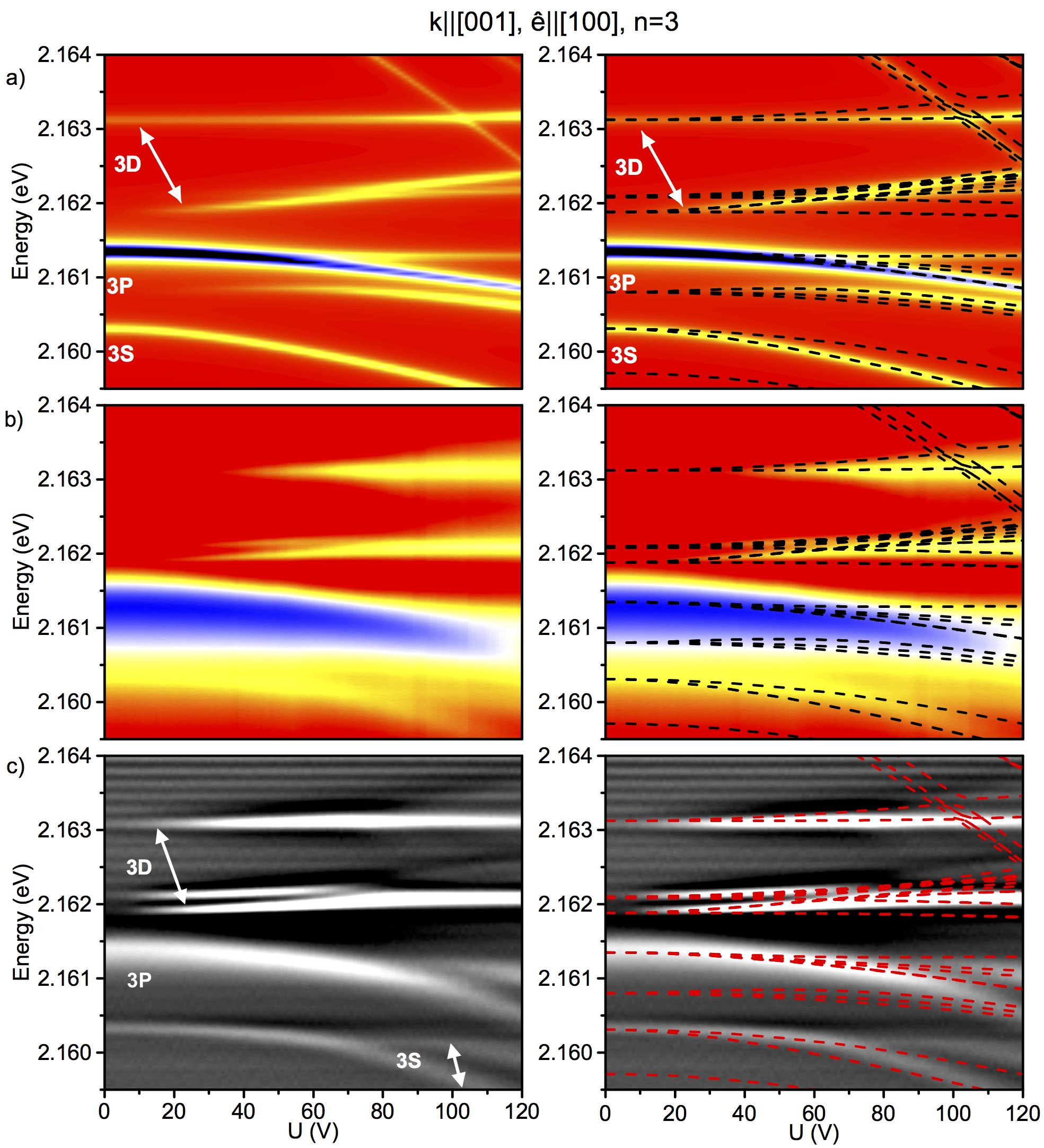}%
\caption{(Color online). Contour plots of transmission spectra of the $n=3$ exciton for the light wave vector $\bm{k} \parallel$ [001], sample 1, as function of applied voltage. The experimental data as recorded on sample 1 are shown in the mid left panel. To bring out tiny features, the second derivative of these spectra over energy is shown in the bottom left panel. The top left panel shows the calculated spectral lines with the intensity scaled by the oscillator strength so that only the optically active transitions contribute. For comparison, in the top right panel the energy dispersions of all the states in the relevant energy range are displayed by the lines. These lines are also superimposed on the experimental data in the two lower figures in the right column. Note that the weak equidistant features along the energy axis in the two lower panels are artefacts of taking the second derivative.
}
\label{fig:3_001}%
\end{figure*}

The experimental data presented in Sec.~\ref{sec:results} have revealed that the electric field induced changes of the transmission spectra show an increasing degree of complexity for increasing principle quantum number $n$. The combined angular momenta of electron and hole plus that of the exciton envelope lead to a large state multiplicity for a particular $n$. In absence of the field the degeneracy of the multiplet at a given $n$ is lifted by the spin-orbit, exchange, and crystal-field interactions accounted for by the terms $ \mathcal H_0 + \mathcal H_{so} + \mathcal H_{exch} + \mathcal H_{cubic}$ in the Hamiltonian~\eqref{H}. 
Further, the application of the field $\bm E$ lifts many of the remaining degeneracies. Moreover, the symmetry breaking in the crystal in combination with the field application leads to a strong state mixing due to which exciton oscillator strength is shuffled from the states optically active without field to inactive states at $E=0$, so that new lines emerge and more and more of them contribute to the measured spectra.

As outlined above, for allowing a quantitative comparison we focus on the states with $n=3$ to $5$. We start with the $n=3$ exciton, which provides a relatively simple but non-trivial and quite illustrative example of the involved physics and interactions. In this case, the angular momentum of the envelope used as an approximate quantum number is restricted to $l \leq 2$, so that only $S$-, $P$-, and $D$-excitons become involved. Optically active in dipole approximation are the $P$-excitons, while through application of the electric field the $S$- and $D$-excitonic states gain oscillator strength so that they also become observable in the transmission experiment.

\begin{figure}[h]%
\includegraphics[width=\linewidth]{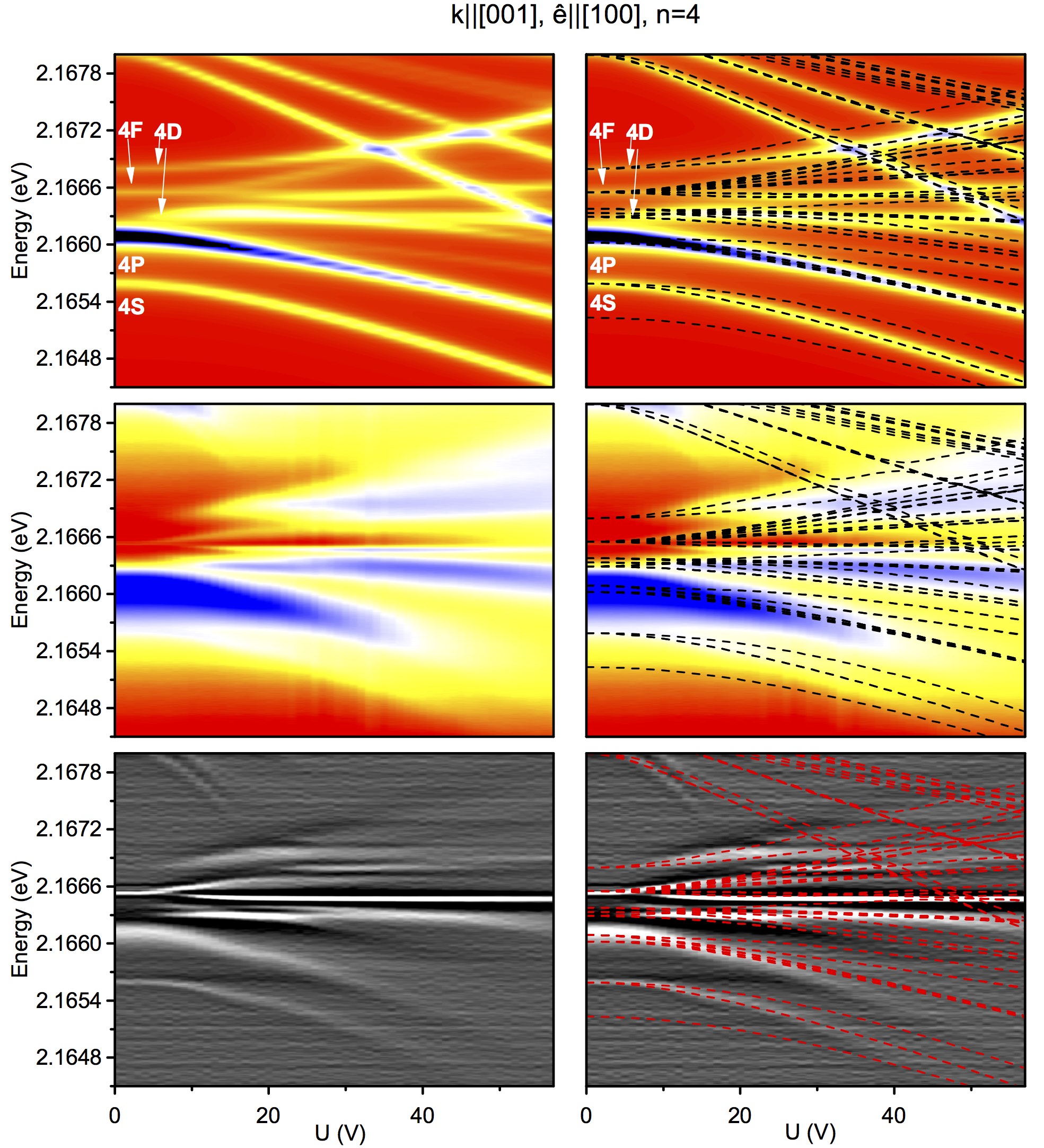}%
\caption{(Color online). Same as Fig.~\ref{fig:3_001}, but for the $n=4$ exciton in sample 1: $\bm {k} \parallel$ [001], $\hat{\bm e} \parallel$ [100].}%
\label{fig:4_001}%
\end{figure}

\begin{figure}[h]%
\includegraphics[width=\linewidth]{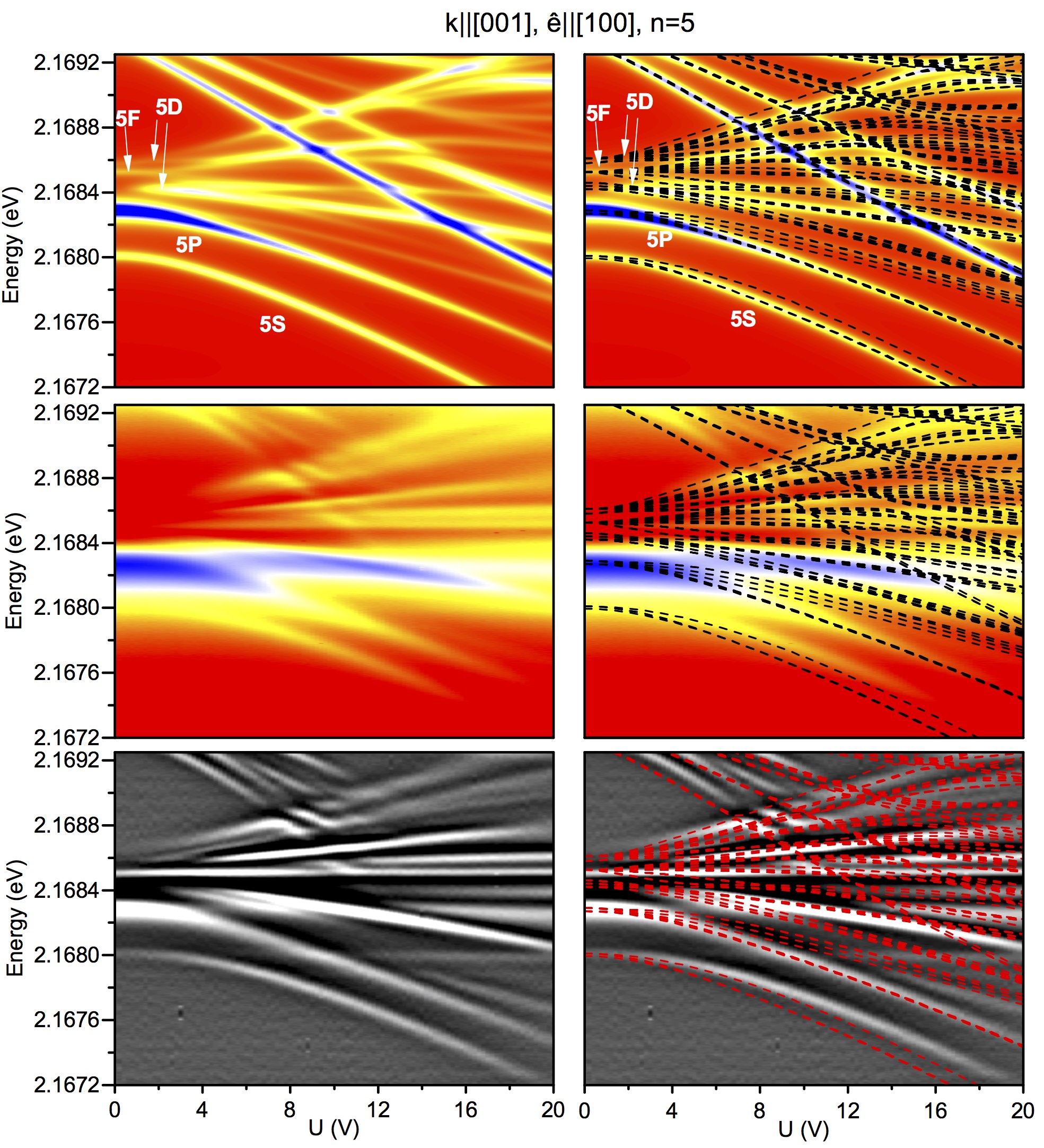}%
\caption{(Color online). Same as Fig.~\ref{fig:3_001}, but for the $n=5$ exciton in sample 1: $\bm {k} \parallel$ [001], $\hat{\bm e} \parallel$ [100].}%
\label{fig:5_001}%
\end{figure}

\begin{figure}[t]%
\includegraphics[width=\linewidth]{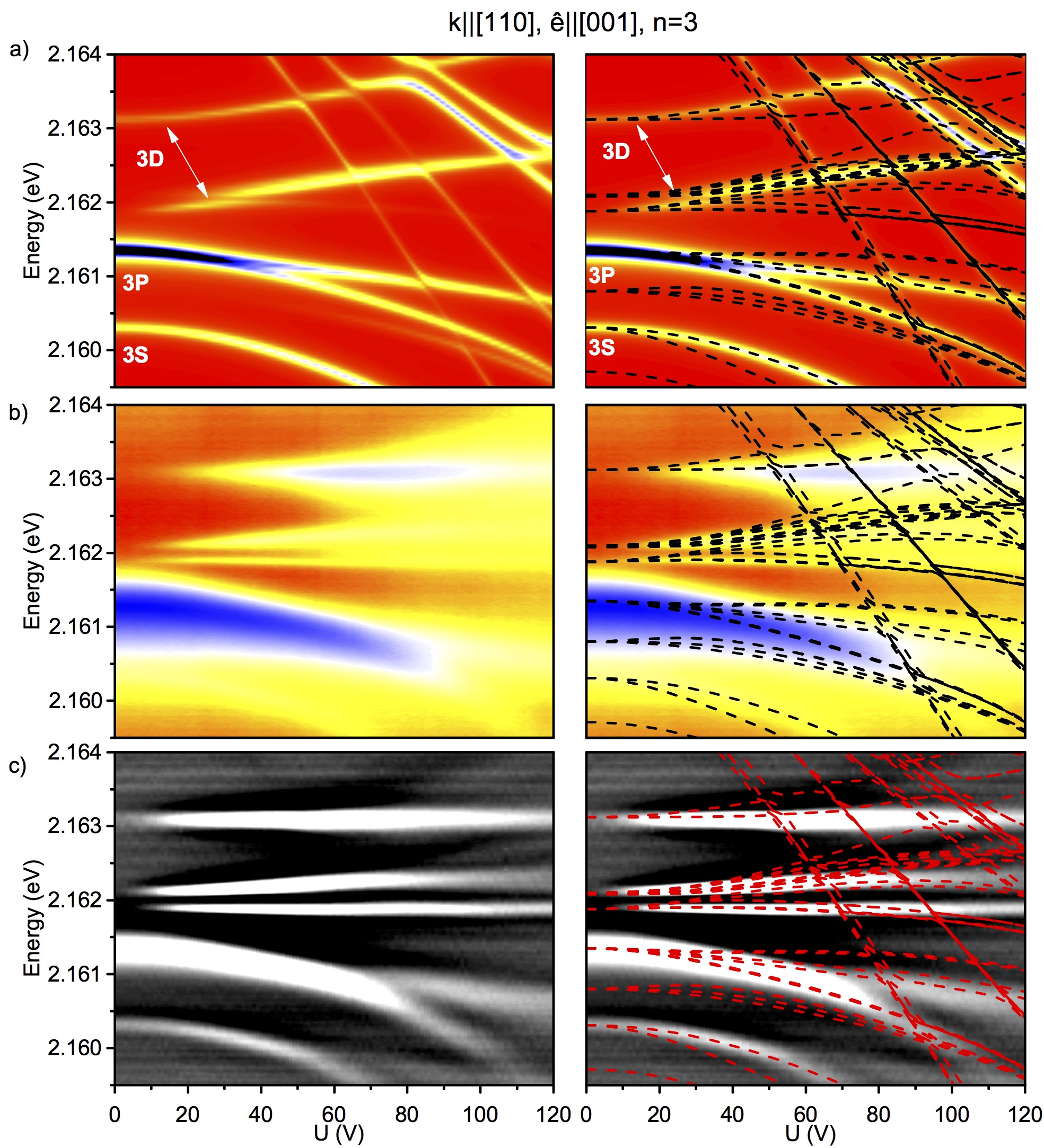}%
\caption{(Color online). Transmission spectra recorded for the $n = 3$ exciton on sample 2 with $\bm{k} \parallel [110]$ and $\hat{\bm e} \parallel [001]$. The contents of the different panels are equivalent to Fig.~\ref{fig:3_001}.}%
\label{fig:3_110}%
\end{figure}

\begin{figure}[t]%
\includegraphics[width=\linewidth]{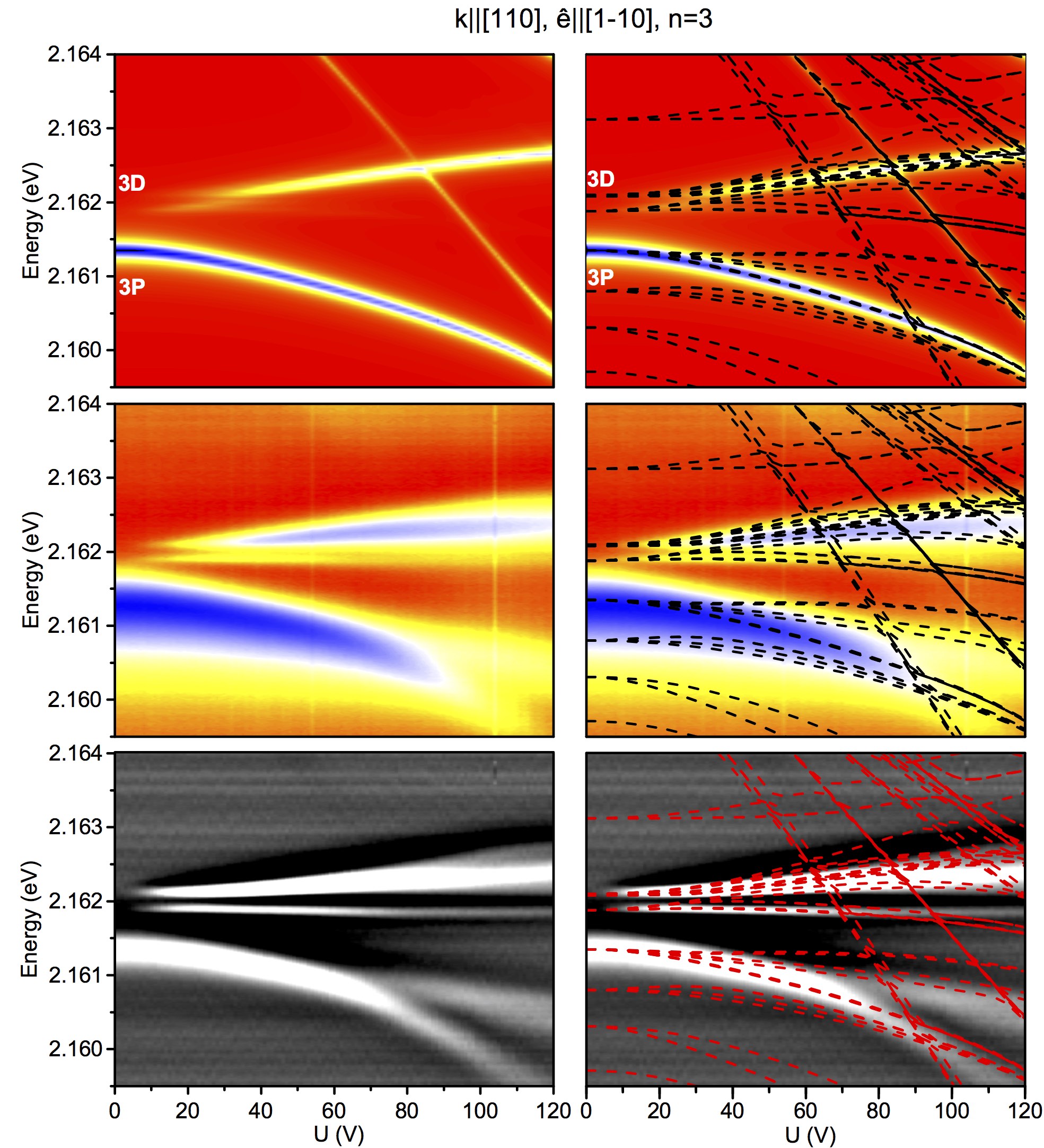}%
\caption{(Color online). Transmission spectra recorded for the $n = 3$ exciton on sample 2 with $\bm{k} \parallel [110]$ and $\hat{\bm e} \parallel [1\bar10]$. The contents of the different panels are equivalent to Fig.~\ref{fig:3_001}.}%
\label{fig:3_110_110}%
\end{figure}

The data for the $n=3$ exciton multiplet are summarized in Figs.~\ref{fig:3_001}, \ref{fig:3_110}, and \ref{fig:3_110_110}. The top left panels of these figures show contour plots of the calculated (after~Eq.~\eqref{abs}) transmission spectra, where only the lines with finite oscillator strength are shown. The color scale is adjusted to correspond to the experimental values. In our numerical calculations we have limited the number of states to those with principle quantum number $n<7$. Inclusion of higher energy states does not change significantly the behavior of lines with the $2\leqslant n \leqslant 5$, which are in the focus of the present work. Let us start the discussion with the case of light propagating along the cubic axis. The middle and bottom left panels in Fig.~\ref{fig:3_001} show the measured transmission spectra of the sample 1, oriented along the [001] direction, on an absolute scale and taking the second derivative, respectively. The light polarization is linear, ${\hat{\bm e}\parallel}$ [100]. At zero field the spectrum is clearly dominated by the $P$-exciton. In accordance with the selection rules this state is identified as the $3P_{3/2}$-exciton, see Tab.~\ref{tab:D}. The second derivative reveals that the $S$-exciton also contributes to the transmission, which we attribute to the quadrupolar transition to this state, Tab.~\ref{tab:Q}. When applying the electric field both states shift to lower energy and the $S$-$P$ splitting increases slightly. The quadratic-in-$E$ Stark shifts for $S$- and $P$-shell excitons are expected from the analogy with Rydberg atoms~\cite{PhysRevA.20.2251}, because these states are already split at $\bm E=0$, unlike in the case of hydrogen where the accidental Coulomb degeneracy of the $n>1$ states results in a linear-in-$E$ Stark effect~\cite{ll3_eng}.
On the high energy side three more features, with two of them close in energy, appear which shift to higher energy as a consequence of repulsion from the $P$-excitons. By energy extrapolation to $\bm E=0$ we assign these features to $D$-excitons. Their oscillator strength increases with increasing electric field, while that of the $P$-exciton drops showing the redistribution of oscillator strength within the multiplet.

As mentioned above, the top left figure shows the theoretical spectra calculated after Eq.~\eqref{abs} (see Sec.~\ref{sec:theory} for details). These spectra are normalized to the same color scale as the experiment. For completeness in the top right panel the energies of all exciton states including those without oscillator strength are shown.  Note that for $n=3$ one has 9 orbital states (1 $S$-exciton, 3 $P$-excitons and 5 $D$-excitons) and 36 states in total with account for the two spin projections each for the electron and for the hole. Similarly, for $n=4$ and $n=5$ (see below) there are 64 and 100 states, respectively. Some of these states are Kramers degenerate and are not split in presence of the $\bm E$ field. The lines corresponding to all states in the system are also included in the two lower plots on the right side, to allow for direct comparison with experiment. In our calculations we have slightly adjusted the positions of the excitonic states at $\bm E=0$ as compared with the microscopic calculations of Refs.~\cite{PhysRevB.23.2731,PhysRevLett.115.027402} in order to reach agreement for $\bm E=0$. Otherwise, minor uncertainties in the states positions do not allow to trace clearly the trends with increasing electric field. Polariton effects are neglected since already for $n=3$ the microscopic calculations without these effects provide good agreement with experiment~\cite{PhysRevB.23.2731,PhysRevLett.115.027402}. For higher states the polariton effects should play an even smaller role because of smaller oscillator strength. Moreover, we have introduced a scaling factor to rescale the applied voltage $U$ to the field $E$ because of the screening issue discussed in Sec.~\ref{sec:samples}. The parameters of the calculations are presented in the Appendix~\ref{sec:appendix}. 

In general, the agreement is convincing. We note already at this point that it is important to include the states of higher $n$ into these calculations because only in this case the field dispersion of the $n$=3 exciton levels is reproduced well due to the interaction between the multiplets. For example, without interaction with higher states, especially with the $n=4$ multiplet, the energies of the $3D$-states increase too much with the increase of electric field and do not match the experiment. Both field induced shifts and changes of oscillator strength are in good agreement in experiment and theory. For the $P$-exciton one expects a doublet splitting with a strong and a weak component, as also observed, which correspond to the $P_{3/2,\pm 3/2}$ and $P_{3/2,\pm 1/2}$ exciton states. For the $S$-exciton a single line is expected from the calculations, while we see at high fields a splitting into a doublet which is observable, however, only when taking the second derivative. We attribute this observation to an electric field that is not perfectly aligned along the [001] direction. Despite the efforts in preparing the sample holder there might be surface charges etc., which, in addition to the complex screening, lead to a slight tilt of the effective electric field from the targeted direction. The $D$-excitons are considerably spread out in energy. Note that also for them we find weak indications at zero electric field, in particular in the second derivative of the experimental data and most prominently for the high energy $D$-state. The exchange interaction between electron and hole in the exciton mixes $S$-shell and $D$-shell excitons, Eq.~\eqref{exch:sd}, providing oscillator strength for the $D$-excitons even at ${\bm E}=0$. As expected, the quadrupole-allowed transition to the $D$-exciton has a much lower oscillator strength than the dipole-allowed transitions and the quadrupolar transitions to the $S$-exciton.

Before addressing the dependence of the spectra on the light propagation direction and polarization, let us turn to the $n=4$  and $n=5$ excitons in the same experimental configuration of sample~1. The panels in Fig.~\ref{fig:4_001} and Fig.~\ref{fig:5_001} provide for these excitons the  information analogous to that in Fig.~\ref{fig:3_001}. From the experiment we find that the low energy flank of the $P$-exciton looks similar to the one for the $n=3$ exciton. Note, however, the distinctly different electric field scales at the abscissa. Similar energy shifts as in the $n=3$ case can be achieved for roughly four times smaller electric fields. This is a consequence of the polarizability of the exciton increasing strongly with increasing $n$~\cite{ll3_eng}.

\begin{figure}
\includegraphics[width=\linewidth]{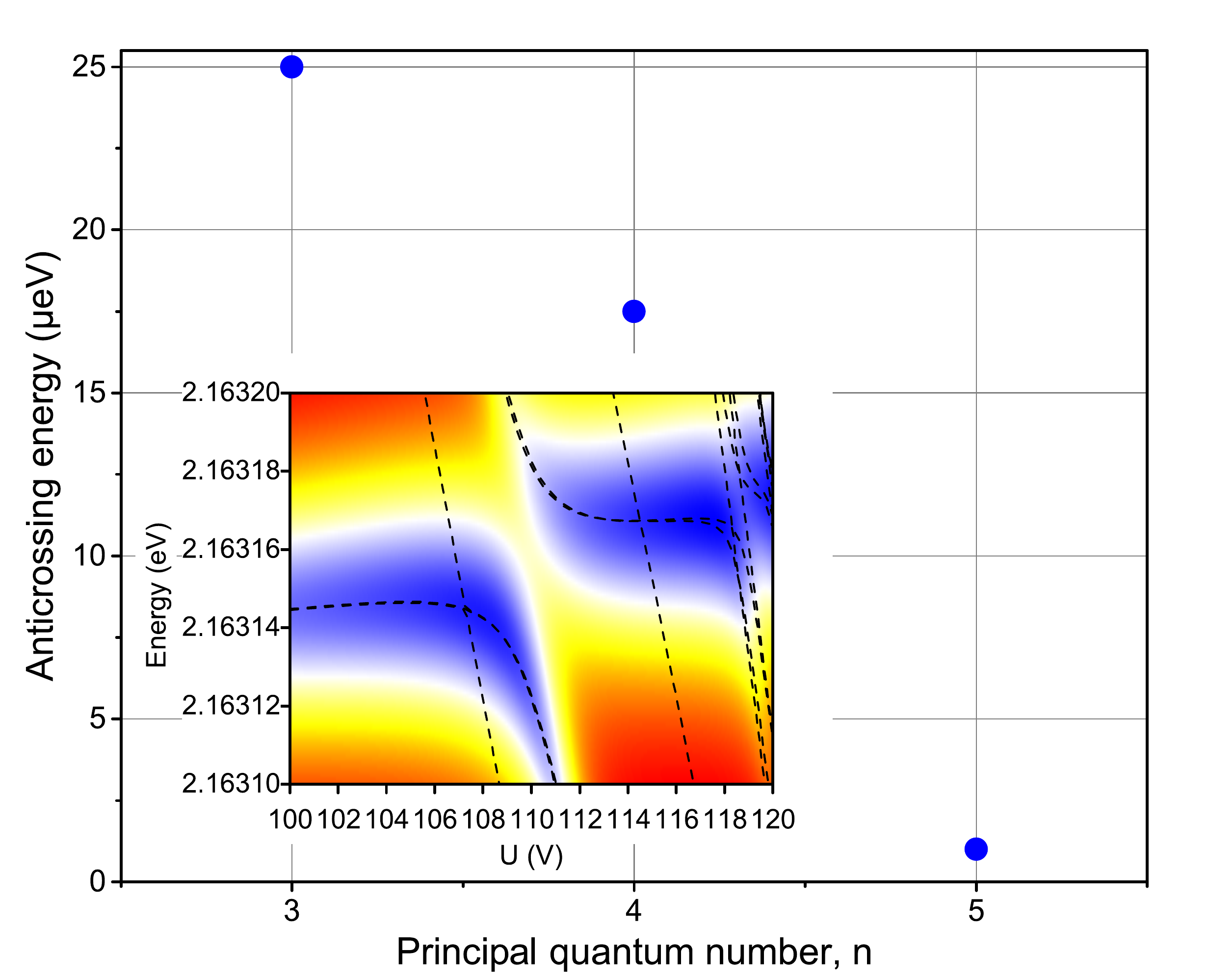}
\caption{(Color online). Splitting energies between states at the first anticrossing of the $n$ and $n+1$ multiplets. Inset shows the calculated spectrum for $n=3$ in the vinicity of the anticrossing with the $n=4$ state descending from higher energies ($\bm k\parallel [001]$, $\hat{\bm e} \parallel [100]$). The dotted lines give the level positions.}
\label{fig:anti}
\end{figure}

\begin{figure}[t]%
\includegraphics[width=\linewidth]{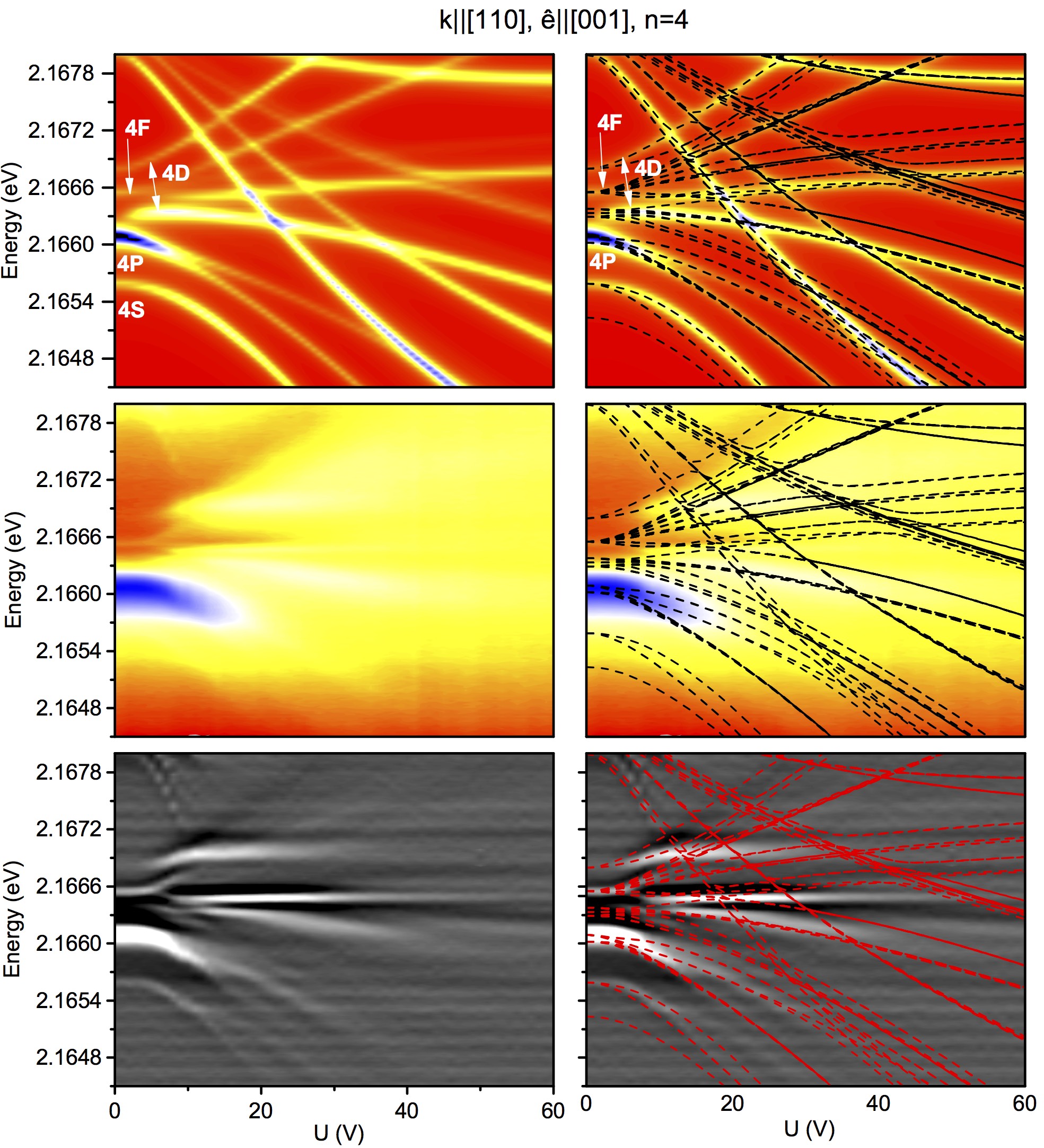}%
\caption{(Color online). Same as Fig.~\ref{fig:3_110}, but for the $n = 4$ exciton in sample 2: $\bm{k} \parallel [110]$ and $\hat{\bm e} \parallel [001]$.}%
\label{fig:4_110}%
\end{figure}

On the high energy side the splitting pattern becomes much more complicated due to the  $F$-excitons, three of which are observed already at zero field, see Fig.~\ref{fig:FG-exc} and Ref.~\cite{PhysRevLett.115.027402}. Theoretically, the interplay of $F$- and $P$-excitons in the presence of electric field has been addressed in Ref.~\cite{Ziemkiewicz}. Further $D$-excitons (as well as $G$-excitons for $n=5$) are activated by the electric field induced state mixing. Additionally, the oscillator strengths of active $P$- and $F$-shell excitons get redistributed in higher orders of $\bm E$, leading to an enhanced complexity of the spectra in this energy range. From the experiment we also find indications for an anticrossing between the levels slightly below 2.167 eV. This is confirmed by the calculations which generally show good agreement with the experiment.

Here the importance of accounting for the exciton states arising from the multiplets with higher principle quantum numbers, i.e., $n = 5$ for the $n=4$ excitons (or $n=6$ for the $n=5$ exciton multiplet) becomes evident, as those states whose energies become lowered by the electric field approach already at moderate fields the $n=4$ levels whose energy simultaneously increases. While the resonances of these levels are only indicated in the measured spectra at low fields they can be well traced in theory. The calculations predict that the optically active levels coming in resonance first from different $n$ weakly anticross each other. The calculation results in Fig.~\ref{fig:anti} show, that for the 
states coming into resonance at lower electric fields and presented here the avoided crossings energies range from $\sim 20$~$\mu$eV to $\sim 1$~$\mu$eV, decreasing strongly with increasing $n$. Hence, the anticrossings are not resolved in the experiment due to broadening of the lines. 

The emergence of anticrossings is an indication that the system may become chaotic for higher principle quantum numbers. This is in accord with recent studies of the yellow Cu$_2$O excitons in  longitudinal magnetic field, where it was shown that the motion stays regular up to $n$ = 6, while for higher $n$ the system turns chaotic in magnetic field as evidenced by suppression of level crossings in favor of anticrossings~\cite{Aszmann:2016aa}. The predominance of crossings in the measured spectra shows that the system behavior is regular in the studied field and energy range. A detailed study of the crossing/anticrossing behavior in electric field is beyond the scope of the present work.

\begin{figure}[t]%
\includegraphics[width=\linewidth]{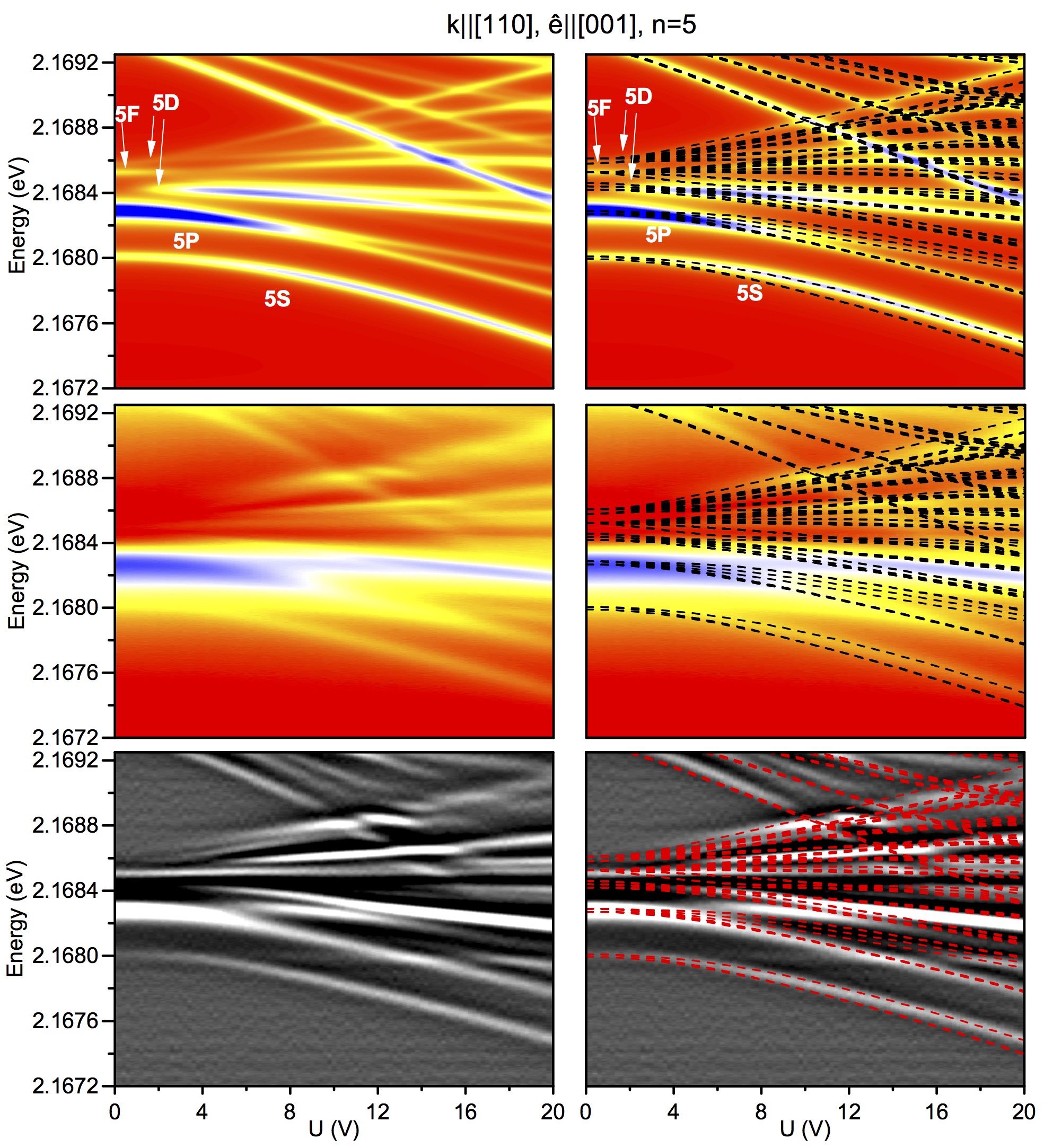}%
\caption{(Color online). Same as Fig.~\ref{fig:3_110}, but for the $n = 5$ exciton in sample 2: $\bm{k} \parallel [110]$ and $\hat{\bm e} \parallel [001]$.}%
\label{fig:5_110}%
\end{figure}

Let us address now the dependence of the transmission spectra on the sample orientation and polarization configuration. To that end we consider the results obtained on the sample 2 with [110] crystal orientation, first with the light linearly polarized along the [001] direction and, subsequently, for $\hat{\bm e} \parallel [1\bar 10]$. For the geometry with $\bm k\parallel [110]$ and $\hat{\bm e} \parallel [001]$ the appearance of the spectra for $n=3$, 4 and 5 in Figs.~\ref{fig:3_110} --- \ref{fig:5_110} is quite similar to the results measured for sample 1. Again, multiple crossings and weak anticrossings between $n=3$ and $n=4$ as well as between $n=4$ and $n=5$ shell states are seen. Particularly noteworthy is the large number of states resulting from the level splitting in electric field as seen prominently in Fig.~\ref{fig:4_110} for the $n = 4$ exciton, mid right panel. Note, however, that many of these states are contributed by the $n = 5$ exciton. 

\begin{figure}[h]%
\includegraphics[width=\linewidth]{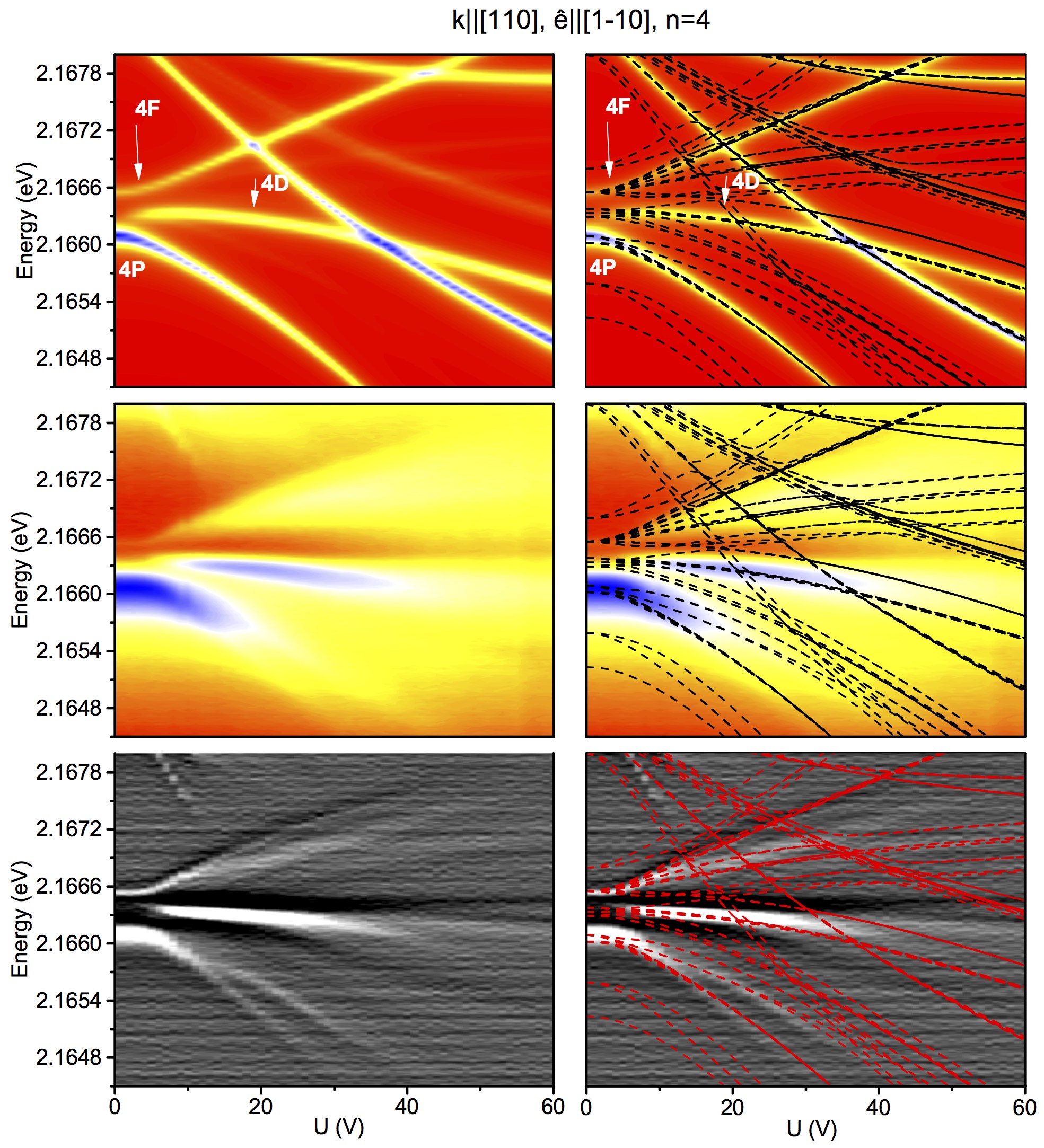}%
\caption{(Color online). Same as Fig.~\ref{fig:3_110_110}, but for the $n = 4$ exciton in sample 2: $\bm{k} \parallel [110]$ and $\hat{\bm e} \parallel [1\bar10]$.}%
\label{fig:4_110_110}%
\end{figure}

\begin{figure}[h]%
\includegraphics[width=\linewidth]{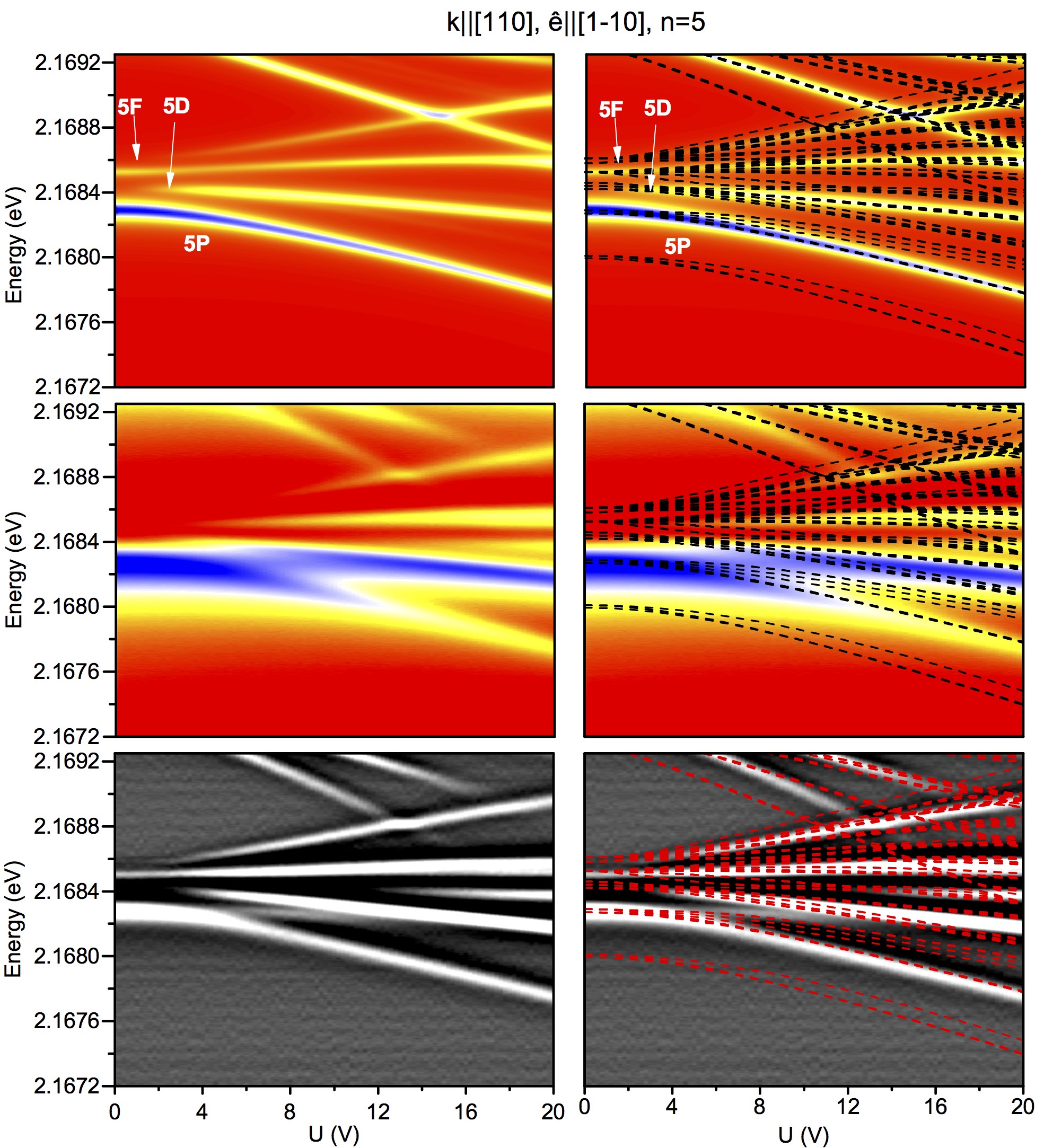}%
\caption{(Color online). Same as Fig.~\ref{fig:3_110_110}, but for the $n = 5$ exciton in sample 2: $\bm{k} \parallel [110]$ and $\hat{\bm e} \parallel [1\bar10]$.}%
\label{fig:5_110_110}%
\end{figure}

From symmetry arguments we expect differences in the transmission spectra at $\bm k\parallel [110]$ for the two orthogonal polarizations: $\hat{\bm e} \parallel [001]$ and $\hat{\bm e}\parallel [1\bar 10]$. Indeed, the transmission spectra of the $n=3$, $4$, and 5 excitons in sample 2 when the light is polarized along the $[1\bar10]$ axis, presented in Figs.~\ref{fig:3_110_110} --- \ref{fig:5_110_110}, are totally different from the results above. Here the calculated spectra show a much smaller number of optically active lines and less diverse splittings when the external electric field is applied.
This simplification arises from the absence of quadrupole allowed transitions in this case, see Sec.~\ref{sec:selection:110} for details. In particular, the $S$-excitons become dark in this polarization configuration. As a result, only those states which get mixed with $P$-excitons (for $n=3,4$) or $P$- and $F$-excitons ($n=5$) appear in the spectra. These predictions are clearly confirmed in the experiment. The recorded spectra show very few more lines than calculated which we attribute again to a slightly inhomogeneous electric field that is not perfectly parallel to the light propagation direction.

For the states with the principle quantum number $n=3$ the main lines in the spectra are associated with the $P$- and $D$-excitons. The most visible line from the $P$-exciton is shifting smoothly towards lower energies.  With increase in electric field, the $D$-excitons start to contribute strongly. Approximately in the same field range the splitting between the $P_{3/2,\pm 3/2}$-excitons and $P_{3/2,\pm 1/2}$-excitons which gain oscillator strength in second order of $\bm E$ appears in the spectrum.  One can see that the doublet of the $D_{5/2}$ states around $2.162$~eV (split by the cubic anisotropy terms, Eq.~\eqref{cubic}) is observed in all three experimental configurations, Figs.~\ref{fig:3_001}, \ref{fig:3_110}, and \ref{fig:3_110_110}, while the upper $D_{3/2}$ state at about $2.163$~eV is visible only for light polarized along one of the cubic axes. Indeed, as the symmetry analysis shows the $D_{3/2}$ state is strongly mixed with the $S$-excitons, Eq.~\eqref{exch:sd}, and becomes active due to quadrupolar effects only.

For the $n = 4$ and $5$ excitons, the transmission tends to remain rather simple, except of the appearance of the $F$-excitons, which are already seen in $E=0$ and are important in the redistribution of the oscillator strength. The interplay of $F$- and $G$-excitons for the $n=5$ multiplet is illustrated in Fig.~\ref{fig:FG-exc}(b). Overall, the splitting between the lowest and highest in energy lines in the multiplets for $n=4$, $5$ increases stronger with increasing $E$ as compared with the $n=3$ excitons. As seen from Fig.~\ref{fig:4_110_110} in this energy range an $n=4$ exciton state comes into resonance with $n=5$ states and we find in experiment in this case crossings confirming that the motion stays regular. We emphasize that a large number of states is involved in the $n=4$ and $n=5$ multiplets, out of which only very few gain oscillator strength. This is a specific consequence of the combination of the [110] light propagation direction and the chosen light polarization $[1\bar10]$.

Overall, the cubic $O_h$ symmetry of the cuprous oxide crystal gives rise to (i) the fine structure of states with high angular momenta, such as the $D_{5/2}$- or $F$-excitons, and (ii) to the modification of the selection rules for optical transitions depending on the light propagation and polarization geometry. In the experiments reported here the effects of cubic symmetry are manifested in the drastic modification of the selection rules, and in particular by activation/deactivation of the quadrupolar transitions depending on the polarization configuration. Additionally, the fine structure of the $D_{5/2}$-state is prominently visible, especially for the $n=3$ excitons.

\section{Conclusion}\label{sec:concl}

In conclusion, we have studied the yellow exciton series in cuprous oxide subject to a longitudinal electric field. The symmetry of the exciton wave function, determined by the electron and hole and the envelope wave function, leads to a large multiplicity of states for a particular principal quantum number. The symmetry reduction in the crystal compared to the hydrogen case becomes enhanced by the electric field application and does not allow one to describe the effects within the simple analogy between the exciton in a semiconductor and the hydrogen atom. As a consequence of the reduced symmetry, the electric field redistributes  oscillator strength between the optically active and forbidden states at $E = 0$, resulting in the activation of initially dark states. We have demonstrated that depending on the crystal orientation relative to the optical axis and the light polarization direction, the transmission spectra differ strongly. This, at first glance, surprising observation of the anisotropy of the optical properties of a cubic crystal, arises from the interplay of dipole-active and quadrupolar-active transitions as well as from the field-induced symmetry reduction. These effects are well described by model calculations taking into account the symmetry of the system and the microscopic details of the optical transitions. 

\acknowledgements

We acknowledge the support of this work by the Deutsche Forschungsgemeinschaft and the Russian Foundation of Basic Research in the frame of the ICRC TRR 160, RFBR-DFG project 15-52-12012, RFBR grants 14-02-00624 (M.M.G), 14-02-01223 (M.A.S), and Russian Federation Government Grant No. 14.Z50.31.0021 (leading scientist M. Bayer). M.M.G. is grateful to the Dynasty Foundation for partial support.

\appendix
\section{Parameters of calculations}\label{sec:appendix}

Tables~\ref{tab:en:orb}, \ref{tab:en:orb:fgh} present the energies of the exciton states used in our calculations which include the spin-orbit interaction. In Tab.~\ref{tab:en:exch} the exchange parameters are presented. In our parametrization the effect is dominated by the $SS$- and $SD$-exchange, Eq.~\eqref{exch:sd}, which provides considerable mixing of the $S$- and $D$-excitons in qualitative agreement with Ref.~\cite{PhysRevB.23.2731}. In order to reproduce the doublets arising from the $D_{5/2}$-states we have included cubic anisotropy terms for the $D$-excitons only, Eq.~\eqref{cubic}. The corresponding parameters are summarized in Tab.~\ref{tab:en:cubic}.

The parameters presented in the tables are obtained by optimization of the agreement between calculated and measured spectra. The precise evaluation of the short-range exchange, spin-orbit and cubic anisotropy constants is beyond the scope of the present work and can be carried out along the lines of Refs.~\cite{PhysRevB.23.2731,PhysRevLett.115.027402,PhysRevB.93.195203}.

The parameters relevant for the calculation of the oscillator strengths are related as follows
\[
ka_B\frac{\mathcal Q}{\mathcal D} = 0.2,
\]
where $k=\sqrt{\epsilon_b}E_g/(\hbar c)$ is the photon wavevector inside the crystal at the exciton resonance frequency, $\epsilon_b$ is the background dielectric constant, $a_B$ is the exciton Bohr radius. The damping of the excitonic state $N_x$ includes two contributions, $\Gamma_{N_x}=\Gamma_1 + \Gamma_0(N_x)$, non-radiative $\Gamma_1=25$~$\mu$eV and the radiative-like one $\Gamma_0(N_x) =\gamma|\tilde{M}(N_x,\hat{\bm e})|^2$. Here $\gamma=150$~$\mu$eV and $\tilde{M}(N_x,\hat{\bm e}) = \langle exc|d|0\rangle/|e\mathcal D a_B^{-5/2}|$ is the dimensionless dipole matrix element of the excitonic state. This contribution is introduced in order to model the state dependent broadening and, in particular, to describe the effect that the states with smaller oscillator strength have narrower lines. For $P$-excitons $\Gamma_0(N_x)$ scales as $1/N_x^3$. Similar dependence of $\Gamma_0(N_x)\propto N_x^{-3}$ is expected for the phonon-induced contributions~\cite{Kazimierczuk:2014yq}. The electric field-induced broadening is disregarded.

    \begin{table}[h]
\caption{Energies (in eV) of the $S$- to $D$-orbital states with account for spin-orbit coupling but neglecting the exchange interaction and cubic anisotropy terms. The energies for the $n=1,2$ excitons are not presented since the calculated spectra are not sensitive to the precise values of these parameters. Hereafter, in the calculations energies close to that in Ref.~\cite{PhysRevB.23.2731} were used. }\label{tab:en:orb}
     \begin{ruledtabular}
     \begin{tabular}{c|c|c|c|c|c}
$n$ & $E(S_{1/2})$ & $E(P_{3/2})$ & $E(P_{1/2})$ & $E(D_{3/2})$ & $E(D_{5/2})$\\
\hline
3 & 2.1607 & 2.16135 & 2.1608 & 2.1622 &2.16249\\
4& 2.16577 & 2.16609 & 2.16602 & 2.1664 & 2.1664\\
5 & 2.16808 & 2.16829 & 2.16827 & 2.16847 & 2.16848\\
6\footnote{Optimization of the $n=6$ exciton energy has not been performed} & 2.1693 & 2.1694 & 2.16939 & 2.16955 & 2.16949
     \end{tabular}
 \end{ruledtabular}
     \end{table}

    \begin{table}[h]
\caption{Energies (in eV) of the $F$- to $H$-orbital states. The spin-orbit and cubic anisotropy effects are neglected for these states.}\label{tab:en:orb:fgh}
     \begin{ruledtabular}
     \begin{tabular}{c|c|c|c|c|c|c}
$n$ & $E(F_{5/2})$ & $E(F_{7/2})$ & $E(G_{7/2})$ & $E(G_{9/2})$ & $E(H_{9/2})$ & $E(H_{11/2})$\\ \hline
4 & 2.16655 & 2.16655 & - & -& - & - \\
5 & 2.16853 & 2.16853 & 2.16861  & 2.16861  & - & -\\
6
&  2.1696 & 2.1696 & 2.16965 & 2.16965 & 2.1697 & 2.1697
     \end{tabular}
 \end{ruledtabular}
     \end{table}
 
     \begin{table}[h]
\caption{Exchange parameters (in $\mu$eV) describing the splitting of the $S$-excitons, $D$-excitons, and the mixing of the $S$- and $D$-excitons.}\label{tab:en:exch}
     \begin{ruledtabular}
     \begin{tabular}{c|c|c|c}
$n$ & $J_n^S$ & $J_{n,3/2}^D$ & $J_{n}^{SD}$ \\ \hline
3 & 1320 & $-160$ & 1228.5 \\
4 & 718.2 & $-30$ & 552\\
5 & 108 & $-14$ & 209.7\\
6
& 10.8  & $-1.4$ & 21
  \end{tabular}
 \end{ruledtabular}
     \end{table}
  
\begin{table}[h]
\caption{Cubic anisotropy contribution (in $\mu$eV), Eq.~\eqref{cubic}, for the $D_{5/2}$-excitons.}\label{tab:en:cubic}
     \begin{ruledtabular}
     \begin{tabular}{c||c|c|c|c}
$n$ & 3 & 4 & 5 & 6 \\ \hline
$\delta_{cubic,nd}$ & $-220$ & $-40$ & $-20$ & 0
  \end{tabular}
 \end{ruledtabular}
     \end{table}

%


\begin{thebibliography}{99}
\bibitem{remote:kp}It can be derived in the $\bm k\cdot \bm p$ perturbation theory taking into account the mixing with odd parity remote bands, e.g., $\Gamma_7^-$ or $\Gamma_8^-$ and the dipole-allowed transitions between the remote and conduction/valence bands~\cite{PhysRev.188.1294}.
\bibitem{note:refl} The analysis of Eqs.~(19) -- (22) of Ref.~\cite{Ziemkiewicz} shows that, if exciton contribution to the reflectivity is disregarded, $n_1 \approx \sqrt{\epsilon_b}$ and $n_2 \ll n_1$, in which case $R\approx [(1-\sqrt{\epsilon_b})/ (1+\sqrt{\epsilon_b})$ is almost frequency independent and $T \approx (1-\alpha L) \times 16 \epsilon_b/(1+\sqrt{\epsilon_b})^2$. Here $\epsilon_b$ is real the background dielectric constant and $\alpha$ is the absorption coefficient related with excitons. Hence, in this approximation the transmission $T$ and abosrption coefficient $\alpha$ are linearly related.
\end{thebibliography}
\end{document}